\documentclass[twocol]{ametsocV6.1}

\usepackage{amsmath,amsfonts,amssymb,bm}
\usepackage{mathptmx}%{times}
\usepackage{newtxtext}
\usepackage{newtxmath}
\newcommand{\pder}[2][]{\frac{\partial#1}{\partial#2}}
\newcommand{\Dder}[2][]{\frac{D#1}{D#2}}

\title{Stratospheric Modulation of the MJO through Cirrus Cloud Feedbacks \\
{\color{red} Submitted for peer review.}}

\authors{Jonathan Lin\aff{a} \correspondingauthor{Jonathan Lin, jzlin@mit.edu}
Kerry Emanuel, \aff{a}}

\affiliation{\aff{a} {Lorenz Center, Department of Earth, Atmospheric, and Planetary Sciences,\\Massachusetts Institute of Technology, Cambridge, Massachusetts}}

\abstract{Recent observations have indicated significant modulation of the Madden Julian Oscillation (MJO) by the phase of the stratospheric Quasi-Biennial Oscillation (QBO) during boreal winter. Composites of the MJO show that upper tropospheric ice cloud fraction and water vapor anomalies are generally collocated, and that an eastward tilt with height in cloud fraction exists. Through radiative transfer calculations, it is shown that ice clouds have a stronger tropospheric radiative forcing than do water vapor anomalies, highlighting the importance of incorporating upper tropospheric/lower stratospheric processes into simple models of the MJO. The coupled troposphere-stratosphere linear model previously developed by the authors is extended by including a mean wind in the stratosphere and a prognostic equation for cirrus clouds, which are forced dynamically and allowed to modulate tropospheric radiative cooling, similar to the effect of tropospheric water vapor in previous formulations. Under these modifications, the model still produces a slow, eastward propagating mode that resembles the MJO. The sign of zonal mean wind in the stratosphere is shown to control both the upward wave propagation and tropospheric vertical structure of the mode. Under varying stratospheric wind and interactive cirrus cloud radiation, the MJO-like mode has weaker growth rates under stratospheric westerlies than easterlies, consistent with the observed MJO-QBO relationship. These results are directly attributable to an enhanced barotropic mode under QBO easterlies. It is also shown that differential zonal advection of cirrus clouds leads to weaker growth rates under stratospheric westerlies than easterlies. Implications and limitations of the linear theory are discussed.}

\begin{document}

%% Necessary!
\maketitle

%%%%%%%%%%%%%%%%%%%%%%%%%%%%%%%%%%%%%%%%%%%%%%%%%%%%%%%%%%%%%%%%%%%%%
% SIGNIFICANCE STATEMENT/CAPSULE SUMMARY
%%%%%%%%%%%%%%%%%%%%%%%%%%%%%%%%%%%%%%%%%%%%%%%%%%%%%%%%%%%%%%%%%%%%%
%
% If you are including an optional significance statement for a journal article or a required capsule summary for BAMS
% (see www.ametsoc.org/ams/index.cfm/publications/authors/journal-and-bams-authors/formatting-and-manuscript-components for details),
% please apply the necessary command as shown below:
%
% Significance Statement (all journals except BAMS)
%
\statement Recent observations have shown that the strength of the Madden Julian Oscillation (MJO), a global-scale envelope of wind and rain that slowly moves eastward in the tropics and dominates global-weather variations on time scales of around a month, is strongly influenced by the direction of the winds in the lower stratosphere, the layer of the atmosphere that lies above where weather occurs. So far, modeling studies have been unable to reproduce this connection in global climate models. The purpose of this study is to investigate the mechanisms through which the stratosphere can modulate the MJO, by using simple theoretical models. In particular, we point to the role that ice clouds high in the atmosphere play in influencing the MJO.
%	 Enter significance statement here, no more than 120 words. See \url{www.ametsoc.org/index.cfm/ams/publications/author-information/significance-statements/} for details.
%
%% Capsule (BAMS only)
%%
%\capsule
%       Enter BAMS capsule here, no more than 30 words. See \url{www.ametsoc.org/index.cfm/ams/publications/author-information/formatting-and-manuscript-components/#capsule} for details.
%
%% * * If using twocol mode, you will need to use the commands "twocolsig" and "twocolcapsule" in place of "sig" and "capsule"
%%      to ensure that the text box correctly spans across both columns.
%

%%%%%%%%%%%%%%%%%%%%%%%%%%%%%%%%%%%%%%%%%%%%%%%%%%%%%%%%%%%%%%%%%%%%%
% MAIN BODY OF PAPER
%%%%%%%%%%%%%%%%%%%%%%%%%%%%%%%%%%%%%%%%%%%%%%%%%%%%%%%%%%%%%%%%%%%%%
%

\section{Introduction}
The MJO is a distinct, eastward propagating, planetary scale oscillation in the tropics that has a period of around 30-90 days, and is the dominant mode of tropical intraseasonal variability \citep{zhang2005madden}. The MJO is also the largest source of seasonal and sub seasonal predictability in the atmosphere \citep{hendon2000medium, vitart2017subseasonal}, and through teleconnections, even plays a significant role in altering extratropical circulations \citep{matthews2004global}. In fact, the MJO has been linked to modulate many aspects of global weather, such as tropical cyclone activity, extreme rainfall and flooding, wildfires, extratropical climate modes, and surface temperatures even in the U.S \citep{zhang2013madden}. As such, furthering our understanding of the MJO is of great societal interest.
%While a widely accepted comprehensive theory on the MJO has yet to be established \citep{zhang2020four}, much progress on its understanding has been made through both theoretical \citep{adames2016mjo, emanuel2020slow, ahmed2021mjo} and numerical modeling efforts \citep{ahn2020mjo}.

Recent studies have uncovered a link between the strength of the MJO and the phase of the Quasi Biennial Oscillation (QBO), a stratospheric mode of variability in which the lower stratospheric zonal winds shift between easterlies and westerlies approximately every 28 months \citep{baldwin2001quasi}. Curiously, the MJO has been observed to be much stronger during the easterly phase of the QBO than the westerly phase of the QBO, but only during boreal winter \citep{yoo2016modulation, son2017stratospheric}. This link has downstream ramifications that are vital; research has shown that the predictability of the MJO is around a week longer during easterly QBO phases than during westerly QBO phases \citep{marshall2017impact}. As a result, sub-seasonal to seasonal forecast models all show enhanced MJO prediction skill during easterly QBO winters \citep{wang2019impact, lim2019influence}. Thus, understanding the physical mechanism through which the QBO can modulate the MJO could help extend the predictability of sub-seasonal forecasts in the tropics, advance modeling of teleconnections between the tropics and extra-tropics, and improve predictions of global climate.

Several mechanisms have been proposed to explain how the mean state of the stratosphere can so strongly influence the strength of a tropospheric phenomenon in the MJO. Since the QBO is associated with vertical wind shear of the zonal wind, thermal wind balance necessitates temperature anomalies in the tropopause transition layer (TTL) \citep{baldwin2001quasi, fueglistaler2009tropical}. One branch of proposed mechanisms contends that during easterly QBO phases, cold anomalies induced by adiabatic cooling destabilize the TTL, invigorating deep convection associated with the MJO \citep{son2017stratospheric, klotzbach2019emerging, abhik2019influence}. However, tropospheric temperature anomalies associated with the QBO are less than 0.5 K in boreal winter \citep{martin2021influence}, and climate models with realistic QBO temperature signals fail to capture the QBO-MJO relationship \citep{martin2021mjo}. Other studies have proposed that the QBO modulates the production of thin cirrus clouds near the tropopause, through mean-state changes in the temperature and stratification in the TTL \citep{sakaeda2020unique}.

One relatively unexplored area is how the QBO can modulate wave propagation into the stratosphere, since the extent to which tropospheric waves can propagate upwards into the stratosphere can be strongly dependent on the sign of the zonal wind in the stratosphere \citep{charney1961propagation, andrews1987middle}. As \citet{charney1961propagation} showed, the upward propagation of tropospheric extratropical Rossby waves is non-linearly dependent on the sign and strength of zonal flow: under easterly or strong westerly flow, Rossby waves are trapped in the troposphere. A similar effect holds in the tropics, where Rossby waves can only propagate upwards in regions of westerly or weak easterly flows \citep{andrews1987middle}. Equatorial Kelvin waves exhibit the opposite dependence, where they can only propagate in regions of easterly or weak westerly winds. Indeed, there is evidence in re-analysis data that Rossby waves are trapped in the troposphere during easterly phases of the QBO, and leak into the stratosphere during westerly phases of the QBO; conversely, Kelvin waves have been found to radiate more energy into the stratosphere during easterly lower stratosphere winds \citep{yang2012influence}. Since the MJO projects strongly onto both equatorial Kelvin and Rossby waves \citep{hendon1994life}, it would be prudent to understand how the QBO can modulate the vertical structure of the MJO through controls on upward wave propagation. Throughout this study, an upwards propagating wave means one that has upwards wave-energy propagation.

Simple theoretical models \citep{sobel2013moisture, adames2016mjo} and idealized modeling studies \citep{crueger2015effect, khairoutdinov2018intraseasonal} have also suggested that cloud-radiative feedbacks are essential to destabilizing the MJO. Given the importance of the modulation of troposheric radiative cooling by clouds, suggested pathways for how the QBO modulates the MJO have included the modulation of cirrus clouds by the stratosphere \citep{son2017stratospheric, sakaeda2020unique}. There is some evidence that the production efficiency of high clouds may be modulated by the phase of the QBO, at least on interannual \citep{davis2013interannual} and seasonal \citep{tseng2017temperature} timescales, since easterly QBO phases are associated with cold anomalies near the tropopause. However, analyses of observational data from the polar-orbiting CALIPSO satellite \citep{winker2009overview} suggest only small differences in near-tropopause cirrus cloud frequency between easterly and westerly phases of the QBO, though the data are generally too sparse in space and time to provide significant evidence \citep{son2017stratospheric}. Furthermore, the QBO does not seem to significantly modulate the activity of other convectively coupled equatorial waves (CCEWs), which may suggest that modulation of cirrus clouds by the QBO is not a significant process \citep{abhik2019sensitivity}. This, however, could be mitigated by the fact that other CCEWs have a much weaker cloud-radiative feedback than the MJO \citep{sakaeda2020unique}.

Other observational studies have suggested a link between cirrus cloud formation and large-scale vertical motion by upward propagating waves \citep{boehm2000stratospheric}. In fact, analyses of satellite observations of temperature and cirrus clouds show that MJO convection is associated with large-scale Kelvin and Rossby wave activity in the TTL, suggesting that the large-scale ascent associated with these waves produces greater levels of cirrus clouds \citep{virts2014observations}. More importantly, equatorial composites of the MJO show significant anomalies in upper-tropospheric/lower-stratospheric ice cloud fraction collocated with MJO convection, as well as an eastward tilt with height in cloud fraction near the stratosphere \citep{virts2010annual, del2015cloud}. This eastward tilt with height in cloud fraction could be the result of dynamical motion associated with the upward propagating Kelvin wave portion of the MJO. The extent to which upward propagating waves influence the MJO growth rate through modulation cirrus clouds will be explored in this study. Finally, the eastward tilt with height could also be explained by mean westerly advection in the upper troposphere. The effect of upper-tropospheric advection of cirrus clouds on the MJO will also be analyzed in this work.

% zonal winds associated with the QBO during boreal winter are small but non-zero [1-3 m/s] in the TTL \citep{son2017stratospheric}. Differential advection of cirrus clouds by the mean flow could modulate the phase relationship of radiative heating with tropospheric convection, influencing the strength and propagation speed of the MJO.

Given the connection between the QBO and MJO, as well as the possibility for ice clouds high in the troposphere to strongly influence the MJO, it is important to incorporate upper-tropospheric/lower-stratospheric processes into models of the MJO. In general, modeling studies on the MJO-QBO link have been particularly limited, since the MJO is notoriously difficult to simulate correctly in a GCM (general circulation model) \citep{hung2013mjo}. Furthermore, an investigation into the MJO-QBO relationship using a nudged GCM was not successful in replicating the observed relationship between the MJO and QBO \citep{martin2021mjo}.  The primary purpose of this study is to investigate the relationship between the MJO and QBO by using an idealized, linear model that can represent cloud radiative feedbacks and tropospheric energy loss via upward wave propagation. Such a model must have some representation of the MJO and also be coupled to a representation of the stratosphere. \citet{khairoutdinov2018intraseasonal} and \citet{emanuel2020slow} developed a strict quasi-equilibrium tropospheric theoretical model and showed that slow, MJO-like modes appear as solutions when cloud-radiative feedbacks are active. \citet{lin2021effect} extended the linear model by coupling a dry, passive stratosphere to a quasi-equilibrium troposphere, and evaluated the effect of upward wave radiation on equatorial waves, though in the context of a zero-mean zonal wind in the stratosphere. We further extend the work of \citet{lin2021effect} by formulating the model for a non-zero zonal wind in the stratosphere, and include an additional prognostic equation for cirrus clouds, which are allowed to modify the perturbation radiative heating in the troposphere.

The paper is organized as follows. Data used in this study to motivate the linear model are described in section~\ref{sec_data}. Section~\ref{sec_rad_forcing} investigates the role of ice clouds in radiative forcing. Section~\ref{sec_linear_model} formulates the linear model. Section~\ref{sec_linear_solutions} presents the solutions of the linear model under varying cases. The paper concludes with a discussion and summary in section~\ref{sec_summary}.

\section{Data \label{sec_data}}
While this study formulates a theoretical linear model to understand stratospheric influences on the MJO, a few observational data sources are used to facilitate formulation of the linear model. Monthly data regarding zonal wind climatology is taken from ERA5 re-analysis fields developed by the European Center for Medium-Range Weather Forecasts (ECMWF), from 1979-2020 \citep{hersbach2020era5}. These data are used in particular to define the QBO, and to examine tropopause transition layer wind anomalies during different QBO phases. In this study, the QBO is defined using the zonal-mean zonal wind at 50-hPa, averaged over the tropics (10$^\circ$S-10$^\circ$N); the QBO is said to be in its easterly phase (QBOE) when the zonal-mean zonal wind is smaller than -0.5 standard deviations from the mean, and in its westerly phase (QBOW) when the zonal-mean zonal wind is greater than 0.5 standard deviations from the mean, as in \citet{son2017stratospheric}. ERA5 re-analysis is also used to generate temperature and water vapor soundings for use in radiative calculations, as well as to generate composites of the MJO.

The phase and amplitude of the MJO are defined using the monthly-averaged OLR MJO Index (OMI), as defined in \citet{kiladis2014comparison}. The OMI index is defined purely based on satellite observations of outgoing longwave radiation (OLR). This is different from the Realtime-Multivariate MJO (RMM) index, which is defined by the two leading principal components (RMM1 and RMM2) of a combination of the equatorially averaged upper (200-hPa) and lower level (850-hPa) zonal winds, and satellite observations of OLR. The phase of the MJO is defined in the phase space of RMM1 and RMM2, following the convention of \citet{wheeler2004all}, with PC2 of OMI being analogous to RMM1, and -PC1 of OMI being analagous to RMM2 \citep{kiladis2014comparison}. The amplitude of the MJO is defined as the magnitude of the monthly-averaged OMI vector, or ($\sqrt{[\text{OMI1}]^2 + [\text{OMI2}]^2}$). The monthly-averaged OMI vectors (instead of the more typical daily quantities) are used since ice cloud observational data is aggregated monthly to increase sample robustness. This, however, likely led to some noisiness in the Phase 4/-Phase 8 composites.

Observations of OLR are taken from NOAA's Interpolated Outgoing Longwave Radiation dataset \citep{liebmann1996description}. OLR anomalies are deseasonalized using monthly averages calculated over the time period 1974-2021. Observations of ice water content and cloud fractions are taken from 2007-2017 Level 3 cloud occurrence products made by the CALIOP instrument on board the polar orbiting CALIPSO satellite. Level 3 products are gridded and aggregated monthly, with a vertical resolution of 60~m. Cloud fraction anomalies are deseasonalized and accumulated over nonoverlapping boxes of width 10$^\circ$ longitude and 5$^\circ$ latitude.

\section{Cloud radiative feedbacks \label{sec_rad_forcing}}
Observational studies have shown that on intra-seasonal time scales, variations in tropospheric radiative cooling are strongly correlated with variations in clouds \citep{johnson2000rainfall}. Convection moistens the troposphere and gives rise to upper tropospheric clouds, thus reducing tropospheric radiative cooling through the greenhouse effect, as both water vapor and clouds absorb infrared radiation and re-emit it at lower temperatures. In fact, on intra-seasonal time scales, there are high correlations between mid-level entropy anomalies (moisture deficit) and outgoing longwave radiation \citep{bony2005role}. These observations have informed the use of mid-level moisture anomalies to predict fluctuations in tropospheric radiative cooling in theoretical linear models; these closures slow down the propagation speed of equatorial waves \citep{bony2005role} and give rise to a new class of unstable modes that resemble the MJO \citep{khairoutdinov2018intraseasonal, emanuel2020slow}.

\begin{figure*}
 \center \noindent\includegraphics[width=30pc,angle=0]{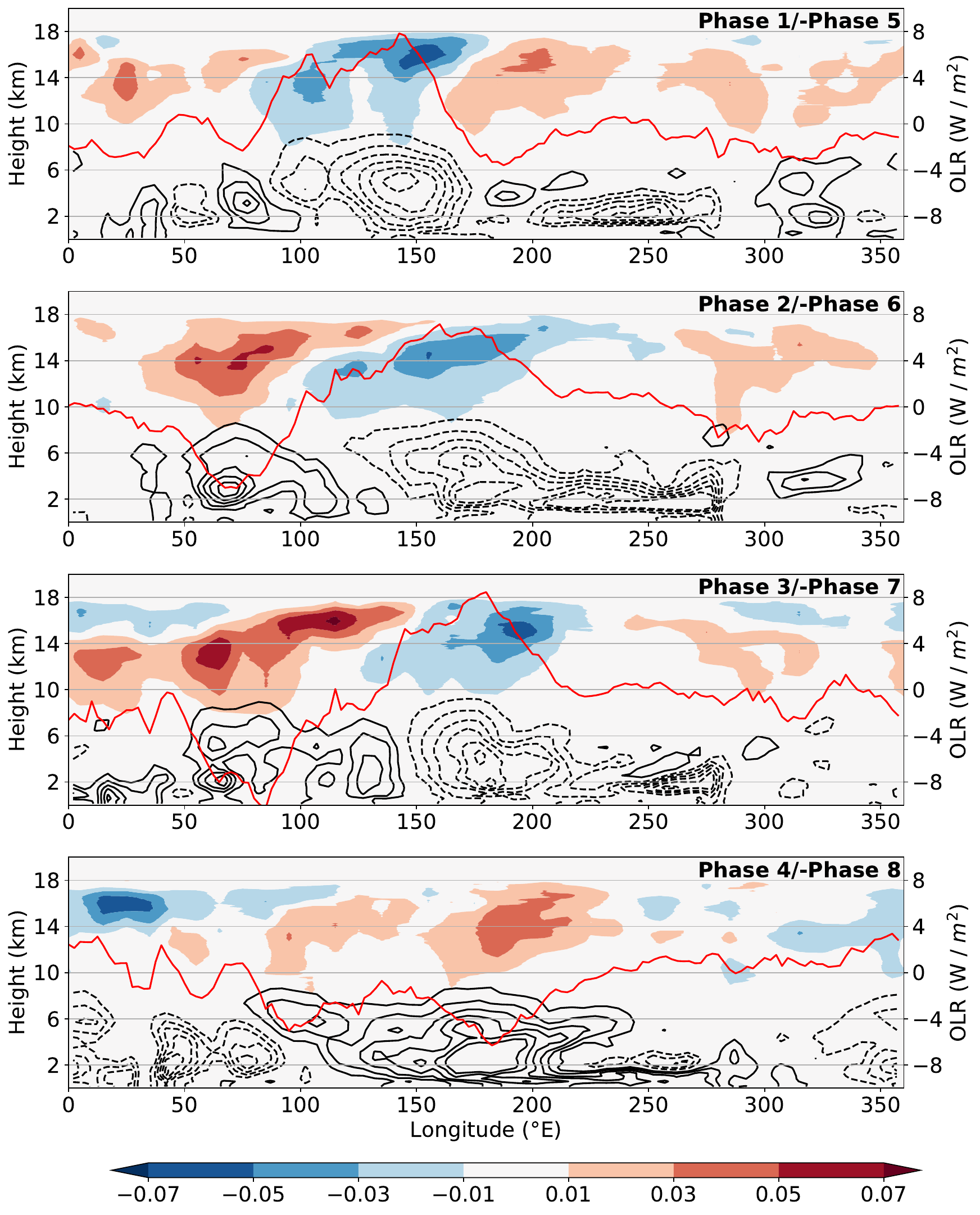}
 \caption{Zonal-vertical cross-sections of the tropical averaged ($5^{\circ}S-5^{\circ}$), monthly anomalies of (colors) ice cloud fraction and (contours) water vapor, aggregated over phases of the MJO, from 10 years (2007-2016) of level 3 CALIOP cloud occurrence profiles and ERA5 re-analysis. Ice cloud samples are deseasonalized and accumulated over boxes of width 10$^{\circ}$ longitude. Phases are determined using monthly RMM index, as defined in \citet{wheeler2004all}. Cloud fraction anomalies are averaged over the indicated phases, but weighted according to MJO amplitude. Contours are solid (dashed) for positive (negative) anomalies. Contour levels start at -0.3 g kg$^{-1}$ with spacings of 0.05 g kg$^{-1}$. OLR anomalies over the composites are overlaid (red line).} \label{fig1}
\end{figure*}

It is instructive to look at the relationship between water vapor, convection, cirrus clouds, and OLR with respect to the MJO. While OLR and lower tropospheric water vapor anomalies are relatively well observed, cirrus clouds are currently only widely observed via polar orbiting satellites, which severely limits the sample size in both space and time. Thus the ensuing analysis should be viewed with this caveat in mind. Figure \ref{fig1} shows tropical averaged (5$^\circ$S-5$^\circ$N) ice cloud fraction (via CALIOP/CALIPSO), water vapor (via ERA5), and OLR (via NOAA) aggregated over combined phases of the MJO, following the methodology of \citet{virts2010annual}. The phases of the MJO are defined following the convention of \citet{wheeler2004all}. A label of ``Phase 1/-Phase 5" aggregates normal anomalies from Phase 1 with anomalies multiplied by -1 from Phase 5, increasing sample size for the mean composites, which are weighted by MJO amplitude. Any further decompositions by QBO phase or season leads to minuscule sample sizes that preclude meaningful analysis.

% Figure 1: Weighted composite of MJO-associated ice clouds, with different phases

The eastward progression of the MJO is quite evident as one moves downward from the top to bottom panels of Figure \ref{fig1}, though the Phase 4/-Phase 8 aggregate has the noisiest signal. In general, mid-level water vapor anomalies are collocated with ice cloud fraction anomalies, and OLR is reduced in areas with more ice clouds and low-level water vapor, and vice-versa. Furthermore, most notably in the Phase 2/-Phase 6 and Phase 3/-Phase 7 aggregates, there are significant ice cloud anomalies (nearly 7\% in cloud fraction) present near or at the climatological lapse-rate tropopause (15-16 km), as well as in the upper portion of the TTL ($\approx$ 17 km). While the ice cloud anomalies are more or less collocated with lower tropospheric water vapor anomalies until around 13 km, there seems to be a substantial eastward tilt with height past $\approx$ 14 km. For instance, in Phase 2, low-level water vapor anomalies are centered around 70$^\circ$ E, but significant positive cirrus cloud anomalies extend eastward by nearly 30$^\circ$ longitude, and as high up as 17 km. This eastward tilt with height in the cirrus cloud fraction is also quite evident in similar MJO-composites of cirrus clouds shown in \citet{virts2010annual}. \citet{del2015cloud} also observe a significant eastward tilt with height in cloud frequency associated with the MJO, and found that MJO-associated cirrus cloud anomalies peak quite high in the atmosphere (15 - 16 km). This eastward tilt with height may be significant for the MJO, especially if cirrus clouds play a large role in modulating tropospheric radiative cooling. However, it is not immediately obvious which of (1) mid-level water vapor or (2) upper-tropospheric ice clouds is more important to tropospheric radiative forcing through the greenhouse effect. Since both quantities are highly correlated with OLR (see Figure \ref{fa0}), linear regression would not be able uncover the relative effects of each quantity on OLR.

\begin{figure*}
 \noindent\includegraphics[width=39pc,angle=0]{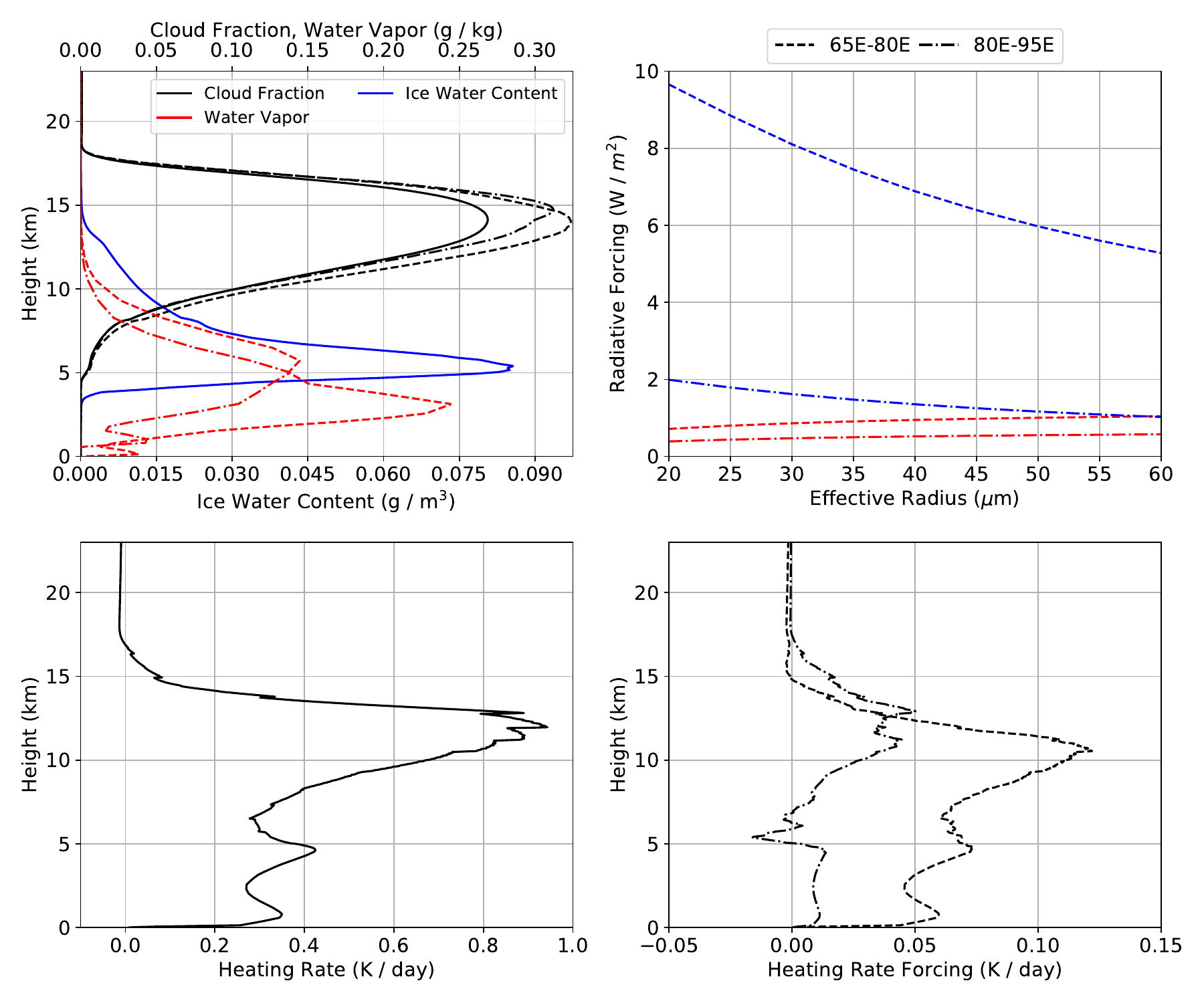}
 \caption{(Top-left) Vertical profiles of (black) cloud fraction, (red) water vapor, and (blue) in-cloud ice water content, where solid lines for cloud fraction and water vapor are tropical-averaged mean-state profiles used in radiative calculations. Vertical profiles of anomalous cloud fraction and water vapor, averaged over (dashed) 65-80$^\circ$E and (dot-dashed) 80-95$^\circ$E in the Phase 2/-Phase 6 MJO-weighted composites, are also shown. (Top-right) TOA radiative forcing, as defined in text, of (blue) ice clouds and (red) water vapor anomalies, for varying effective radius. (Bottom-left) Vertical profile of heating rate from ice-clouds, in the mean-state. (Bottom-right) Anomalous heating rate from anomalous ice clouds in the (dashed) 65-80$^\circ$E and (dot-dashed) 80-95$^\circ$E composites.} \label{fig2}
\end{figure*}

To address these issues, we use the Rapid Radiative Transfer Model (RRTM), supplied with typical anomalies of mid-level moisture and ice clouds associated with the MJO, to obtain an order-of-magnitude estimate of the radiative forcings of these two quantities \citep{mlawer1997radiative}. Specifically, we use the longwave radiative transfer code of RRTMG\_LW, the computationally accelerated version of RRTM developed for GCMs.  Ice-cloud and liquid clouds are treated separately in these calculations. In this study, we do not consider the radiative effect of lower-tropospheric liquid clouds, since the fractional area occupied by liquid water clouds is much smaller than ice clouds (verifiable using the CALIPSO/CALIOP cloud fraction data). The parameterization of ice-cloud optical properties follows that developed in \citet{ebert1992parameterization}. The model's mean-state is tropical averaged ($10^{\circ}S-10^{\circ}N$) temperature and water vapor soundings, calculated from ERA5 re-analysis during the same period over which the ice-cloud observations are available, along with 400 ppm $CO_2$, and 1.7 ppm $CH_4$. The mean-state cloud fraction and cloud ice water content are estimated from CALIPSO data by averaging the cloud fraction and monthly median ice water content, respectively, over ($10^{\circ}S-10^{\circ}N$). Figure \ref{fig2}, top-left, shows the mean-state cloud fraction and ice water content profiles derived from the CALIPSO data. To generate typical anomalies of water vapor and ice clouds associated with the MJO, we focus on the convective region of the Phase 2/-Phase 6 weighted composite (Figure \ref{fig1}). Vertical profiles of anomalous cloud fraction and water vapor, averaged over (dashed) 65-80$^\circ$E and (dot-dashed) 80-95$^\circ$E, are shown in Figure \ref{fig2}. The reasoning for this delineation is as follows: ice clouds are collocated with convection around 65-80$^\circ$E, coinciding with the largest negative OLR anomaly, whereas an eastward tilt with height and upward shift of the peak ice cloud anomaly occurs in the 80-95$^\circ$E region. This can be seen from Figure \ref{fig1}, and from the cloud fraction vertical profiles in Figure \ref{fig2}. The cloud fraction anomalies in the 80-95$^\circ$E region, however, are smaller in magnitude than those in the 65-80$^\circ$E reigon. Finally, we do not modify in-cloud ice water content, since the ice water content of upper tropospheric clouds is very small, and primarily a function of temperature.

% Figure 2: Radiative forcing of ice clouds and water vapor

The top-of-atmosphere (TOA) radiative forcing of water vapor and upper tropospheric ice clouds is defined as the difference in OLR between the mean-state, and a state with the MJO-associated water vapor or cloud fraction imposed. Figure \ref{fig2}, right, shows that the TOA radiative forcing of ice-clouds is nearly an order of magnitude larger than that of mid-level water vapor in the main convective region (65-80$^\circ$E), though there is some sensitivity to the effective radius of ice clouds. Note, the median effective radius of ice clouds in the tropics is $\approx$ 35 $\mu m$ \citep{hong2015characteristics, hong2016assessing}. At this effective radius, the RRTM-calculated total radiative forcing of the water vapor and ice-cloud anomalies is around 8 W m$^{-2}$, which is in agreement with the aggregated OLR anomaly in this region, as shown in Figure \ref{fig1}. The radiative forcing of ice clouds is also around 90\% of the total radiative forcing. In the region east of the main convective region (80-95$^\circ$E), the ice-cloud radiative forcing is, as expected, much smaller, though still non-negligible. At an effective radius of 35 $\mu m$, the radiative forcing of ice clouds is only around 75\% of the total radiative forcing. The vertical profile of heating rate by ice-clouds in the mean-state, calculated by subtracting clear-sky heating rates from all-sky heating rates, is shown in Figure \ref{fig2}, bottom-left, and shows a substantial tropospheric warming effect by ice-clouds. These results are generally consistent with those of \citet{hong2016assessing}, except for the discrepancy that the present study suggests a much smaller stratospheric radiative cooling response to upper-tropospheric ice clouds. This discrepancy can be explained by differences and uncertainties in ice-cloud optical depth measurements at high altitudes. For optically thin clouds in the tropics, the CALIPSO product (used in this study) has significantly smaller ice water content retrievals than DARDAR, another cloud retrieval product that derives ice cloud parameters from both radar and lidar measurements \citep{delanoe2010combined, hong2016assessing}. In fact, the signal-to-noise ratio of upper troposphere-lower stratosphere ice cloud net long-wave radiative heating rates is smaller than 1 (noise here, defined as the difference between the net radiative heating rates between the CALIPSO and DARDAR products). The reader is referred to the appendix of \citet{hong2016assessing} for more details.

While the relative magnitude of radiative forcing between ice clouds and water vapor shows some sensitivity to the effective radius and vertical distribution of ice clouds, these radiative transfer calculations suggest that cirrus clouds play a dominant role in modulating tropospheric radiative heating on spatial and temporal scales similar to that of the MJO. Thus, modulation of the forcing of ice clouds in the TTL may be a pathway through which the QBO controls the MJO. In addition, despite the sparsity of satellite observations as shown in Figure \ref{fig1}, there is evidence that MJO-associated cirrus clouds near the equator exhibit an eastward tilt with height in the TTL. Modulation of the eastward tilt of these cirrus clouds by the stratosphere could influence the phase relationship of radiative heating with temperature in the troposphere.  Both of these mechanisms will be explored in a coupled troposphere-stratosphere linear model in the ensuing section.

%We propose two mechanisms through which this could occur: (1) dynamical modulation of cirrus clouds by upward propagating waves, and (2) advection of cirrus clouds by the background wind. On the first mechanism, anomalous vertical motion from upward propagating Kelvin waves, which have eastward tilts with height, could be responsible for near-equator TTL cirrus cloud anomalies. In regards to the second mechanism, it is also possible for cirrus clouds to be advected westward by upper tropospheric westerlies. The effect of these two mechanisms will be explored in a coupled troposphere-stratosphere linear model in the ensuing section.

\section{Linear model \label{sec_linear_model}}
In this section, we extend the coupled troposphere-stratosphere model formulated in \citet{lin2021effect}, by imposing non-zero stratospheric mean zonal wind, as well as incorporating the effects of cirrus clouds on radiative cooling by including a prognostic equation for cirrus clouds.

\subsection{Tropospheric equations}
Here, we summarize the tropospheric equations of the linear model formulated in \citet{lin2021effect}, taking the case where surface friction is set to zero ($F = 0$). In particular, \citet{lin2021effect} removed the rigid-lid in the tropospheric equations, which allows the barotropic mode to be excited in a linear model:
\begin{align}
    \pder[u_0]{t} &= -\pder[\phi_0]{x} + y v_0 \label{eq_uBT} \\
    \frac{1}{\delta_x} \pder[v_0]{t} &= -\pder[\phi_0]{y} - y u_0 \label{eq_vBT} \\
    \pder[u_1]{t} &= \pder[s]{x} + y v_1  \label{eq_uBC} \\
    \frac{1}{\delta_x} \pder[v_1]{t} &= \pder[s]{y} - y u_1  \label{eq_vBC} \\
    \pder[s]{t} &= (1 + C) s_m - w - \alpha u_b - \chi s \label{eq_s} \\
    \gamma \pder[s_m]{t} &= -D s - \alpha u_b - G w + C s_m \label{eq_sm}
\end{align}
where $u_0$ and $v_0$ are the barotropic zonal and meridional winds, $u_1$ and $v_1$ are the baroclinic zonal and meridional winds, $\phi_0$ is the barotropic geopotential, $\delta_x$ is a non-dimensional coefficient representing the magnitude of zonal geostrophy, $s$ is the saturation moist entropy, $s_m$ is the characteristic mid-level moist entropy of the free troposphere, $w$ is the bulk tropospheric vertical velocity, $u_b$ is the boundary layer zonal wind (equal to $u_0 + u_1$), $\chi$ is a non-dimensional entropy damping coefficient, $\gamma$ is a non-dimensional tropospheric entropy time scale, $D$ is a non-dimensional entropy damping coefficient, $\alpha$ is the wind-induced surface heat exchange (WISHE) feedback parameter, and $G$ is the gross moist stability. The equations are completed with mass continuity in the troposphere in pressure coordinates. Note that in these equations, the radiative heating perturbation is parameterized as $Q = C s_m$, which is evident in Equations (\ref{eq_s}) and (\ref{eq_sm}). This is important, since \citet{emanuel2020slow} showed that when $C$ is non-zero and large enough, slow propagating modes that are MJO-like appear as the fastest growing modes. The reader is referred to \citet{lin2021effect} for additional details in derivation, non-dimensionalization, and interpretation of Equations (\ref{eq_uBT}) - (\ref{eq_sm}), which are incomplete without governing equations for the stratosphere. Modifications to the stratospheric equations are discussed in the following sections.

\subsection{Non-zero stratospheric mean wind}
To include non-zero stratospheric mean wind, we must modify the stratospheric equations formulated in \citet{lin2021effect}, which assumed a zero-mean zonal wind. The non-dimensional vertical velocity at the tropopause can be inferred by integrating the mass continuity equation upwards from the surface.
\begin{equation}
    \omega(y, \hat{p}_t) = \pder[u_0]{x} + \pder[v_0]{y}
\end{equation}
where $u_0$ and $v_0$ are the barotropic velocities $\omega$ is the pressure vertical velocity, and $\hat{p}_t$ is the non-dimensional tropopause pressure. Note the baroclinic velocities do not enter here since by definition, the vertical velocity associated with the baroclinic mode vanishes at the tropopause.

Next, we assume that the mean zonal wind in the stratosphere is non-zero and varies in the vertical:
\begin{equation}
    u_{s}(x, y, z^*, t) = \overline{U}_s(z^*) + u_s^\prime(x, y, z^*, t)
\end{equation}
where $z^*$ is the log-pressure vertical coordinate. For simplicity, the time scale of the mean stratopheric wind is assumed to be much longer than that of the tropospheric wave, and thus the mean wind is assumed to be constant in time. After dropping primes for perturbation quantities and using the additional non-dimensionalization of the mean zonal wind:
\begin{equation}
	\overline{U}_s \rightarrow \beta L_y^2 U_s
\end{equation}
the linearized, non-dimensional horizontal momentum equations in log-pressure coordinates of the stratosphere are:
\begin{align}
    \pder[u_s]{t} + U_s \pder[u_s]{x} + \Gamma_m w^*_s \pder[U_s]{z^*} &= -\pder[\phi_s]{x} + y v \label{eq_duSdt} \\
    \frac{1}{\delta_x} \Big( \pder[v_s]{t} + U_s \pder[v_s]{x} \Big) &= -\pder[\phi_s]{y} - y u \label{eq_dvSdt}
\end{align}
where subscripts of $s$ indicate stratospheric variables, $w^*_s$ is the log-pressure perturbation vertical velocity, $\Gamma_m$ (defined in the Appendix) is a non-dimensional parameter corresponding to the strength of vertical zonal momentum flux, $L_y$ is the meridional length scale defined in \citet{khairoutdinov2018intraseasonal}, and $\beta$ is the meridional gradient of the Coriolis force. Typical peak zonal wind anomalies at 50-hPa are on the order of 20~m~s$^{-1}$ during QBOE and 15~m~s$^{-1}$ during QBOW \citep{baldwin2001quasi}, which yields an approximate range of the non-dimensional mean wind  as $U_s \approx [-0.75, 0.5]$. While non-dimensional scaling suggests that $\Gamma_m \approx 0.5$, it is not immediately clear how large the vertical zonal momentum flux term is in relation to the other terms, since the vertical shear of the QBO can be quite large. For the MJO-like mode explored in this study, we confirmed that under an Earth-like QBO, the vertical zonal momentum flux term is an order of magnitude smaller than the pressure gradient and Coriolis forces, which are largely balancing each other. Thus, the vertical zonal momentum flux can be ignored without significant approximation, and we set $\Gamma_m = 0$ throughout this study. Note that this approximation may not be accurate for faster propagating waves.

The mass continuity equation in the stratosphere is unchanged from \citet{lin2021effect}, though it is provided here for completeness:
\begin{equation}
    \pder[]{x}u_s + \pder[]{y} v_s + \frac{1}{\rho_0} \pder[]{z} \Big( \rho_0 w^*_s \Big) = 0 \label{eq_cont}
\end{equation}
where $z^* = 1$ is defined as the tropopause (lower boundary), and $\rho_0(z^*) = \exp (\alpha_H (1 - z^*))$ is a non-dimensional density that decays with a non-dimensional scale height $\alpha_H$. With the approximation that the vertical wavelength of the mode of interest is much smaller than $\alpha_H$, Equation (\ref{eq_cont}) can be integrated from the lower boundary in $z^*$, to obtain:
\begin{equation}
    w^*_s(y, z^*) = w^*_s(y, z^* = 1) - \int_{z = 1}^{z} \Big[ \Big( i k u(y, z^*) + \pder[]{y} v(y, z^*) \Big) \Big] dz \label{eq_w_strat}
\end{equation}
where $w^*_s(z^* = 1)$ is coupled to the vertical velocity at the tropopause in the troposphere equations and will be calculated from the matching conditions.

The first matching condition is continuity of normal displacement across the interface (in dimensional notation), defined as:
\begin{equation}
    w = \Dder[\eta]{t}
\end{equation}
where $\eta$ is the displacement at the tropopause. While we do assume mean-wind in the stratosphere, for simplicity, we do not assume a jump in mean wind across the tropopause, such that the coupling condition is unmodified from that formulated in \citet{lin2021effect}, and is simply continuity of vertical velocity across the tropopause, or $w_s(\hat{p} = \hat{p}_t) = w_t(\hat{p} = \hat{p}_t)$. The vertical velocity between the two vertical-coordinate systems are matched:
\begin{align}
	w_s(\hat{p} = \hat{p}_t) &= \frac{T_t}{\overline{T}_s} w^*_s(z^* = 1)  \label{eq_wstrat} \\
	w_t(\hat{p} = \hat{p}_t) &= -\frac{R_d T_t}{g H} \frac{\hat{p}_s - \hat{p}_t}{\hat{p}_t} \omega(\hat{p} = \hat{p}_t) \label{eq_wtrop}
\end{align}
where $T_t$ is the tropopause temperature, $\hat{p}_s$ is the non-dimensional surface pressure, $\overline{T}_s$ is the mean temperature in the stratosphere, $R_d$ is the dry gas constant, $g$ is the acceleration of gravity, and $H$ is the tropopause height.

\subsection{Thermal wind balance}
The stratospheric equation set is not yet complete, as we have yet to formulate its mean temperature equation. As is observed in the real-atmosphere, QBO-associated vertical gradients in mean zonal wind must be associated with meridional gradients in temperature, according to thermal wind balance \citep{baldwin2001quasi}.  For an equatorial $\beta$-plane, thermal wind balance is expressed as [see equation 8.2.1 in \citet{andrews1987middle}]:
\begin{equation}
    \pder[u_s]{z} = -\frac{R_d}{\beta H_s y} \pder[T]{y}
\end{equation}
where $H_s$ is the scale height in the stratosphere. Non-dimensionalizing temperature with:
\begin{equation}
    T \rightarrow \frac{\beta^2 L_y^4}{R_d} T
\end{equation}
yields:
\begin{equation}
 \pder[\overline{U}_s]{z} = - \frac{H}{H_s} \frac{1}{y} \pder[\overline{T}]{y}
\end{equation}
for the mean-state field. Non-dimensionalizing similarly in the hydrostatic equation yields:
\begin{equation}
     \pder[\phi_s]{z^*} = \xi T \label{eq_hydrostatic_nd}
\end{equation}
where for a scale height of $H_s = 7$ km and $H = 16$ km, $\xi = \frac{H^2 \beta L_y^2}{H_s a C_k |V|} \approx 70$. Note that the vertical shear in the zonal mean wind associated with the QBO can be large on the equator, yielding peak temperature anomalies of around 4 K \citep{baldwin2001quasi}. This may impact the dynamics of equatorial waves in the stratosphere. To incorporate this into the linear model, we start with the dimensional temperature equation (including hydrostatic balance), which is:
\begin{equation}
    \Big( \pder[]{t} + \Vec{V} \cdot \boldsymbol{\nabla} \Big) \pder[\phi_s]{z^*}  + w^*_s N^2 = 0
\end{equation}
where the squared buoyancy frequency in log-pressure coordinates is:
\begin{equation}
    N^2 \equiv \frac{R_d}{H_s} \Big( \pder[T]{z^*} + \kappa \frac{T}{H_s} \Big)  \label{eq_log_pressure_n2}
\end{equation}
where $\kappa = R_d / c_p \approx 2 / 7$. Note, we assume no radiative relaxation in the stratosphere, as consistent with the radiative transfer calculations shown in Figure \ref{fig2}. It is important to note that while our radiative-transfer calculations indicate minimal impact on stratospheric radiative heating rates by upper-tropospheric ice clouds, large uncertainties in the optical properties of upper-tropospheric/lower-stratospheric ice clouds exist.

QBO contributions to the mean-state stratification are small: a 4 K perturbation over 5 km yields a perturbation buoyancy frequency of $[N^2]^\prime \approx 4~\times~10^{-5}~s^{-2}$, which is more than an order of magnitude smaller than the buoyancy frequency of the stratosphere, $N^2 \approx 5~\times~10^{-4}$~$s^{-2}$. Thus, we approximate $N^2$ as constant. Linearizing the temperature equation under non-zero zonal flow in thermal wind balance yields:
\begin{equation}
    \pder[]{t} \pder[\phi_s^\prime]{z^*} + \overline{U}_s(y, z) \frac{\partial^2 \phi^\prime_s}{\partial x \: \partial z^*} + v^\prime \frac{\partial^2 \overline{\phi}_s}{\partial y \: \partial z^*}  + w^{* \prime}_s N^2  = 0  \label{eq_dphidzdt}
\end{equation}
Non-dimensionalizing Equation (\ref{eq_dphidzdt}) and dropping primes for perturbation quantities yields:
\begin{equation}
     \pder[]{t} \pder[\phi_s]{z^*} + U_s(y, z) \frac{\partial^2 \phi_s}{\partial x \: \partial z^*} + \frac{\xi}{\gamma_v} \pder[\overline{T}]{y} v + w^{*}_s S = 0 \label{eq_strat_temp_nd}
\end{equation}
% \begin{equation}
%      \pder[\phi^\prime_z]{t} + \overline{U}_s i k \phi^\prime_z + \frac{1}{\gamma_v} \pder[\overline{\phi}_z]{y} v^\prime + \frac{H^2}{H_s} \Big( \frac{1}{H} \pder[\overline{T}]{z^*} + \frac{\kappa}{H_s} \overline{T} \Big) w^{*\prime} = 0
% \end{equation}
where $\gamma_v = \frac{H \beta L_y^2 }{C_k |V| a} \approx 30$, such that $\xi / \gamma_v \approx 2.5$, which will be used for the rest of the study. Although meridional temperature gradients associated with the QBO can be large on the equator, the magnitude of the QBO-associated temperature anomalies decay quite quickly away from the equator. The opposite is true for meridional velocities: they can be large off the equator (especially in the Rossby gyres associated with the MJO), but are typically small near the equator. We confirmed in our experiments that for the MJO-like mode that appears in the linear model, the dominant terms in Equation (\ref{eq_strat_temp_nd}) are the zonal and vertical advection terms (the second and fourth terms on the left hand side, respectively).

Next, we implement a wave-radiation condition through Equation (\ref{eq_strat_temp_nd}). As in \citet{lin2021effect}, it is not necessary that $w^*$ goes to zero as $z \rightarrow \infty$: as long as the energy density ($\rho w^{*2}$) goes to zero, then wave energy is forced to be propagating upwards from the troposphere. However, we include strong sponge layers at the top and lateral boundaries of the numerical domain to ensure that the velocities do go to zero at the edges. Integrating Equation (\ref{eq_strat_temp_nd})  from the upper boundary, while ignoring meridional advection and assuming a zero upper boundary condition, yields:
\begin{equation}
    \pder[\phi_s]{t} =  - i k \int^{z}_\infty U_s(y, z) \frac{\partial^2 \phi_s}{\partial x \: \partial z^*} dz^* - \int^{z}_\infty w^*_s S \: dz^* \label{eq_dphidt}
    % - \frac{\xi}{\gamma_v} \int^{z}_\infty \pder[\overline{T}]{y} v^\prime dz^*
\end{equation}
Equation (\ref{eq_dphidt}) is the time-stepping equation for the geopotential in the stratosphere. Finally, $\phi_s(y, z^* = 1)$ couples to the troposphere equations through the second matching condition, continuity of pressure across the interface:
\begin{equation}
    \phi_s(x, y, z^* = 1, t) = \phi(x, y, \hat{p} = \hat{p}_t, t) \label{eq_phi_match}
\end{equation}

% \subsection{Cirrus cloud feedbacks}
% Cold anomalies that are associated with large-scale rising motion of upward propagating waves can induce favorable conditions for formation of cirrus clouds at the tropopause \citep{boehm2000stratospheric}. We attempt to crudely mimic the effect of cirrus cloud formation in the tropical tropopause layer by including a perturbation to the radiative heating based on the temperature perturbation at a level $z$ that is induced by the large-scale vertical motion at that level:
% \begin{equation}
%     Q = C s_m - C_i T_z
% \end{equation}
% where the non-dimensional radiative heating $Q$ is a function of a cloud-radiative feedback $C$ in the troposphere, the non-dimensional cirrus-cloud feedback $C_i$, and $T_z$, the temperature perturbation at a height of $z$. The temperature anomalies are obtained from the perturbation form of Equation (\ref{eq_hydrostatic_nd}). Note we take $C_i > 0$ assuming cirrus clouds exert a warming effect on tropospheric temperature. This means that cold anomalies associated with upward propagating Kelvin and/or Rossby waves will further increase the radiative heating in the troposphere. According to \citep{tseng2017temperature}, the presence of cirrus clouds was most correlated not with the temperature at the tropopause, but around 1 km above the tropopause.

\subsection{Cirrus cloud prognostic equation}
To incorporate the cloud-radiative effects associated with cirrus clouds, we include the linearized, dimensional, water vapor prognostic equation, but only at a single level, $z_c$:
\begin{equation}
    \pder[q_v]{t} + \overline{U}_c \pder[q_v]{x} + w^\prime \pder[\overline{q}_v]{z} = P - L \label{eq_qv_dim}
\end{equation}
where $q_v$ is the water vapor mixing ratio, $\overline{U}_c$ is a mean zonal wind, P is the anomalous production of water vapor, and L is the anomalous loss of water vapor.

First, we comment on the form of this equation. The cirrus cloud prognostic equation only applies at a single level, $z_c$, which represents the level where the presence of cirrus clouds dominates the radiative heating effect in the troposphere. It should be restricted to the upper troposphere, which is where the climatological cloud fraction peaks. On MJO time scales, there is observational evidence of $w$ at a single level being a good predictor of cirrus cloud fraction at that level, though the peak $w$ anomalies often lead the peak ice cloud fraction anomalies. This is described in detail in the Appendix, and serves as justification of the form of Equation (\ref{eq_qv}). Of course, this is an oversimplified view of radiative transfer, and an integrated metric involving $q_v$ is more appropriate to relate to radiative heating perturbations. Our parameterization will serve the purpose of simplified representation of cirrus clouds, and sensitivity tests to $z_c$ are shown in this study. The sign and magnitude of $\overline{U}_c$, as well as the levels at which to parameterize $z_c$, will be discussed in the results section. Furthermore, both the production and loss of water vapor are determined by cloud-microphysical processes, which can be quite complex, especially when considering mixed phase clouds. Here, we take a simplified approach, and assume that production and loss are proportional to $w^\prime$. In general, we expect there to be cloud condensation, growth, and precipitation where there is upwards motion, but this is mostly deduced from qualitative reasoning. Non-dimensionally (and dropping primes) this is:
\begin{equation}
    P - L = C_m w
\end{equation}
where $C_m$ which is a fairly arbitrary coefficient. Non-dimensionalizing Equation (\ref{eq_qv_dim}), with the relations,
\begin{align}
    t &\rightarrow \frac{a}{\beta L_y^2} \\
    x &\rightarrow a x \\
    w^\prime &\rightarrow C_k |\textbf{V}| w \\
    q^\prime_v &\rightarrow \overline{q_i^*} q_v
\end{align}
where $\overline{q_i^*}$ is the mean saturation vapor mixing ratio with respect to ice, yields
\begin{equation}
    \pder[q_v]{t} + U_c \pder[q_v]{x} = (\Upsilon + C_m) w \label{eq_qv}
\end{equation}
where $U_c$ is the non-dimensional advecting wind, and $\Upsilon$ represents the vertical moisture gradient, assumed to be positive (a negative sign is absorbed into $\Upsilon$), and is:
\begin{equation}
     \Upsilon = -C_k |\textbf{V}| \frac{a}{\beta L_y^2 \overline{q^*_i}} \pder[\overline{q_v}]{z} \approx 0.5
\end{equation}
Note that $C_m$, while fairly arbitrary, can be absorbed into the definition of $\Upsilon$, at which point $\Upsilon$ represents the strength of water vapor production. Sensitivity of the results to $\Upsilon$ will be discussed in the concluding section.

In numerical models, a cloud macrophysics closure is required to predict cloud fraction from ice and vapor mixing ratios; these closures typically assume a quadratic relationship between cloud fraction and total water content \citep{gettelman2010global}. Cloud microphysical schemes, which predict growth and loss of cloud ice, are also necessary to predict cloud fraction. In the spirit of simplicity, we will assume that water vapor anomalies can serve as a proxy for cirrus clouds, which in turn modulate the perturbation radiative heating. This assumption is not entirely unfounded: linearization of the ice-only CAM5 cloud macrophysics parameterization, shown in the Appendix, leads to a relation between water vapor anomalies and cloud fraction:
\begin{equation}
    \text{ICF}^\prime = \epsilon_i q_v
\end{equation}
where $\text{ICF}^\prime$ is the perturbation ice cloud fraction, and $\epsilon_i$ represents the production efficiency of ice clouds with respect to water vapor anomalies. Nevertheless, thorough empirical verification of the accuracy of Equation (\ref{eq_qv}) in predicting ice-cloud fraction is necessary, though this will be the subject of future work. Finally, the tropospheric radiative heating in the model is modified to include effects from our proxy for cirrus clouds:
\begin{equation}
    Q = C s_m + C_i [\text{ICF}]^\prime \label{eq_q}
\end{equation}
where $C_i > 0$ is the cirrus cloud feedback parameter. Stratospheric radiative heating is ignored. In general, as $C_i$ increases in magnitude in relation to $C$, more weight is given to the high-cloud parameterization of cloud-radiative feedbacks. Note, the original formulation of cloud-radiative feedbacks in \citet{khairoutdinov2018intraseasonal} is obtained by setting $C_i = 0$. Radiative transfer modeling, discussed in section \ref{sec_rad_forcing}, allows us to constrain $C_i$, especially with respect to $C$. In this study, $\epsilon_i$ is absorbed into $C_i$, which is chosen such that in the vertically integrated entropy equation (Equation \ref{eq_sm}), ice-clouds make up $\approx 80\%$ of the total radiative forcing.

It is important to note that these are simplistic and crude representations of cirrus clouds and the processes that might effect their behavior. While there is evidence in observational data of the modulation of cirrus clouds by upward propagating waves \citep{boehm2000stratospheric, virts2014observations}, the extent to which these processes influence the MJO have yet to be validated with either observations or high-resolution numerical modeling. While Equations (\ref{eq_qv}) and (\ref{eq_q}) do not truly represent the complexity of cirrus cloud formation and cloud microphysics, they are meant to highlight some potential mechanisms that may allow the stratosphere wind to modulate the MJO. It is our intention, in the spirit of simplicity, to understand how each of the modeled processes can affect growth of the MJO.

\subsection{Numerical solutions}
The fully coupled system consists of the tropospheric system [Equations (\ref{eq_uBT}) - (\ref{eq_sm})], the stratospheric system [Equations (\ref{eq_duSdt}), (\ref{eq_dvSdt}), (\ref{eq_w_strat}), (\ref{eq_dphidt})], the matching conditions [Equations (\ref{eq_wstrat}) (\ref{eq_wtrop}), (\ref{eq_phi_match})], and the parameterizations for cirrus cloud feedbacks [Equations (\ref{eq_qv}), (\ref{eq_q})]. Note that we have not assumed anything about the meridional or vertical dependence of $\overline{U}_s$. Unless otherwise noted, once $\overline{U}$ is chosen, the associated $\overline{T}$ is calculated through thermal wind balance.

The linear system is complex and cannot be solved analytically. As described in detail in \citet{lin2021effect}, the system is solved numerically by integrating forwards in time, initializing the troposphere with the rigid-lid solution while the stratosphere is initialized at rest \citep{emanuel2020slow}. The troposphere domain is discretized in $y$, while the stratosphere domain is discretized in $y$ and $z^*$. Linear solutions are assumed to have zonal structure of the form $\exp(ikx)$. Spatial derivatives are numerically approximated with fourth order central differencing, and the system is stepped forward in time using fourth order Runge-Kutta. Since the initial wave is unbalanced, rapid gravity-wave adjustment occurs, requiring the use of dampening mechanisms to eliminate undesirable noise. First, a spectral filter is applied at each time step to eliminate small-scale noise. The spectral filter is described in detail in the appendix of \citet{lin2021effect}. In addition, a strong sponge-layer is imposed along the edges and top of the domain, strongly attenuating reflecting and downward propagating waves. The domain is re-scaled by a constant periodically in time to prevent numerical overflow. After integrating for a long period of time (around 160 Earth days), we isolate the growing mode of interest and infer the complex growth rate and structure of the eigenmode. The complex growth rate is calculated by fitting a line to the log-amplitude and phase of any prognostic variable. The inferred eigenmode and growth rates are then rigorously checked to satisfy the governing equations, boundary conditions, and matching conditions. In general, the numerical solutions are accurate to O($10^{-4}$).

Before proceeding, it is prudent to discuss some of the features of this model. Since there is a mean zonal wind in the stratosphere, a critical layer will develop if the phase speed of the wave equals the mean wind at some level. Linear numerical models are highly unstable in the presence of critical layers. In addition, the behavior of upward propagating waves when encountering critical layers can be highly sensitive to non-linearity and dissipation. Wave-breaking, wave reflection, and a transfer of momentum to the mean flow are quite often associated with critical layers, as is the case for the QBO \citep{lindzen1968theory}. None of these features are represented in this model. However, there are some qualitative aspects of critical layers that linear models can capture, such as attenuation of the wave through the critical layer \citep{booker1967critical}. It is important to note that the very small amount of meridional diffusion imposed in all of the prognostic equations is required for the linear model to maintain numerical stability. This may be because without explicit dissipation, critical layers would be able to form in the stratospheric domain, though this was not thoroughly investigated.

\section{Linear solutions  \label{sec_linear_solutions}}
Since the mechanism through which stratospheric wind can modulate MJO growth is the primary focus of the paper, we will focus our analysis on the eigenmode that most resembles the MJO in the linear model. In light of this, we use the mean-state over the TOGA-COARE Intensive Observing Period (IOP) to inform the non-dimensional parameters of the model \citep{webster1992toga}. All eigenmodes, unless otherwise stated, are computed using the following selection of non-dimensional coefficients: $\alpha = 0.35$, $\chi = 0.2$, $C = 1.25$, $\gamma = 5$, $D = 0.25$, $G = 0.02$, $\Upsilon = 0.5$, $\Gamma_m = 0$, $\delta_x = 30$, $S = 40$. Note, these coefficients are different from those used by \citet{lin2021effect}, which focused on the broad spectrum of equatorial waves. These parameters are described earlier in the text, with the exception of $S$, which represents the magnitude of stratospheric stratification. In all figures, colors shading with blue indicate negative quantities, and red indicate positive quantities.

\subsection{Stratospheric control of vertical structure}
% In this section, show the vertical w structure, and wave patterns in stratosphere
In the coupled troposphere-stratosphere model, both the barotropic and baroclinic modes can be excited. The superposition of the barotropic and baroclinic modes can lead to more complex vertical structures than modes that are purely baroclinic. The barotropic mode is more closely tied to stratospheric dynamics, since, by definition, the barotropic mode is associated with non-zero vertical velocity at the tropopause. As will be shown in this section, stratospheric dynamics can play a significant role in modulating the magnitude of the barotropic mode, as well as the phase relationship between the barotropic and baroclinic modes. Thus, the vertical structure can be heavily modified by the sign of the stratospheric wind.

To first isolate how the sign of the stratospheric wind can impact the vertical structure of the MJO, we look at linear solutions of the MJO-like eigenmode using the original \citet{khairoutdinov2018intraseasonal} radiative cooling parameterization ($C = 1.25$, $C_i = 0$), but under varying stratospheric zonal wind. The simplest, realistic, vertical structure of stratospheric wind is a constant shear in the mean zonal wind, capped at maximum value of $U$; the mean zonal wind increases linearly until it reaches $U$, after which it becomes constant. Mathematically, this is:
\begin{equation}
    U_s(z^*) = \min \Big( \Gamma (z^*-1), U \Big)  \label{eq_us_linear}
\end{equation}
where the reader is reminded that the tropopause is defined at $z^* = 1$. $\Gamma$, which controls the depth of the linear shear layer, is set such that $U_s = U$ at 5-km above the tropopause, which coincides with the depth of the QBO's lowest shear layer. In this study, we assume that the tropopause wind is $U_s(z^* = 1) = 0$. This removes the presence of highly unstable shear instabilities that are the result of discontinuities in mean wind across the tropopause interface.

%% Figure 3 %%
\begin{figure*}
 \noindent\includegraphics[width=39pc,angle=0]{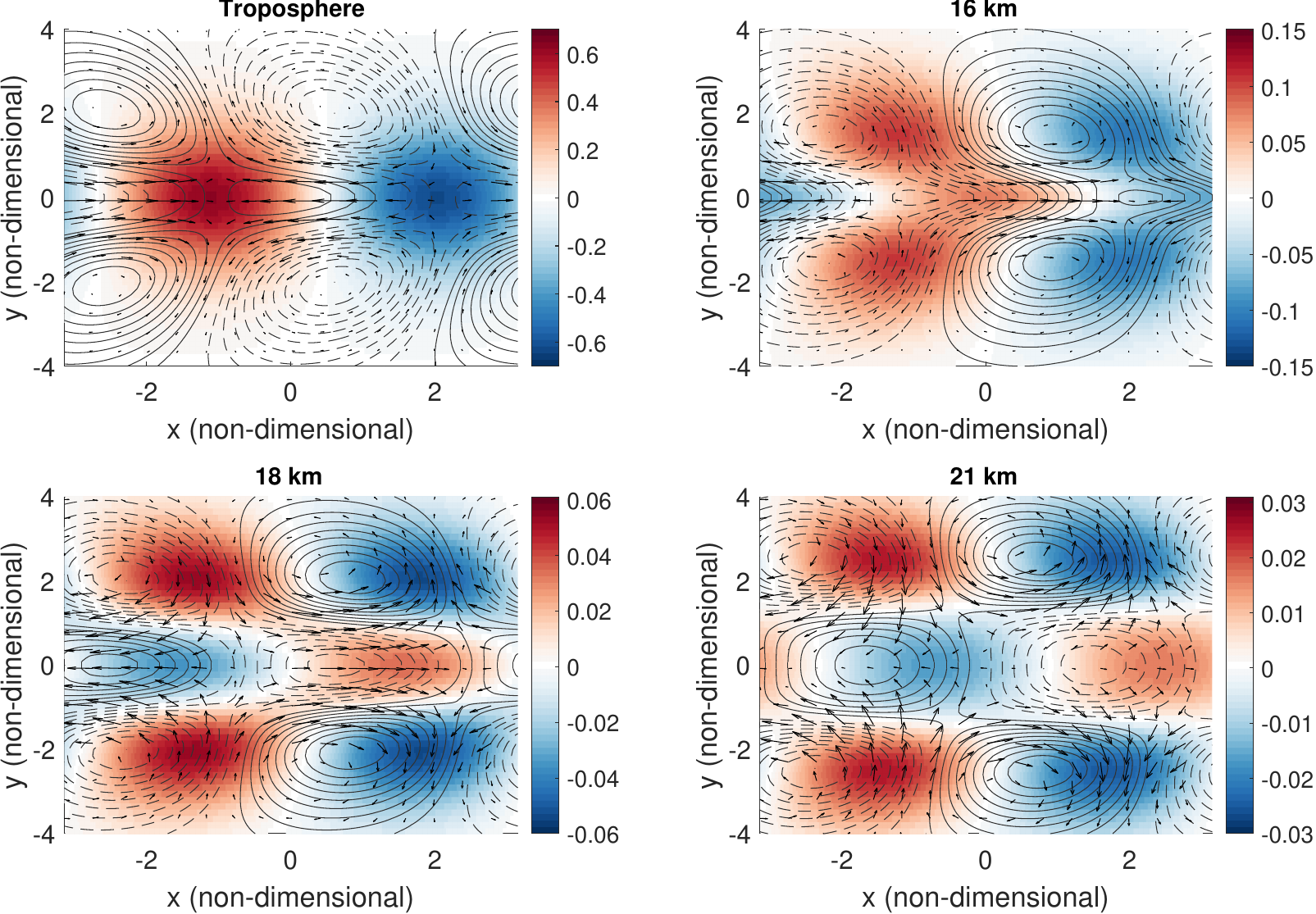}
 \caption{Horizontal cross-sections of the $k = 1$, MJO-like eigenmode at the (top-left) boundary layer, (top-right) tropopause [16-km], (bottom-left) 18-km, and (bottom-right) 21-km, for the case of capped, constant shear in stratospheric wind and $U = -1$ (mean easterlies). Contours indicate the saturation entropy in the troposphere and temperature in the stratosphere, where solid (dashed) lines indicate positive (negative) perturbations. Arrows indicate wind perturbations, and color shadings indicate vertical velocity perturbations at the level indicated (positive for upward), except for the boundary layer cross-section, where color shading indicates mid-level vertical velocity. Non-dimensional parameters selected are $\alpha = 0.35$, $\chi = 0.2$, $\gamma = 5$, $D = 0.25$, $G = 0.02$, $\delta_x = 30$, $S = 40$, $C = 1.25$, and $C_i = 0$.} \label{fig3}
\end{figure*}

Figure \ref{fig3} shows the horizontal summary eigenfunction at different vertical levels of the $k = 1$, eastward propagating, MJO-like mode, under mean easterly winds, $U = -1$. The boundary layer horizontal structure is similar to the MJO-like mode introduced in \citet{emanuel2020slow}; qualitatively, a canonical swallow-tail-like structure is shown, with a Kelvin-wave signature near the equator, lagged and flanked by equatorial Rossby waves. Strong westerly anomalies lag the maximum in vertical velocity on the equator, which is also preceded by strong easterly anomalies. This pattern somewhat resembles the observed MJO, except that westerly maxima are observed to be in phase with the maximum in vertical velocity, a common criticism of WISHE-based theories for MJO destabiliziation \citep{lin1996heating, kiladis2005zonal}. Since the vertical structure of the MJO is dominated by the first baroclinic mode \citep{adames2014three}, the sign of the horizontal wind at the tropopause (Figure \ref{fig3}, top-right) is opposite that in the boundary layer, except for a slight tilt with height in the upper troposphere, as will be discussed in depth later. As we move further up into the stratosphere, there appears to be, at least qualitatively, a separation between the Kelvin-wave component and the Rossby-wave component of the MJO-like mode. The meridional structure of the stratospheric mode seems to be quite complex, as the Kelvin and Rossby wave components seems to be zonally out of phase around halfway into the shear-layer (Figure \ref{fig3}, bottom-left), and at the altitude of the shear-layer (Figure \ref{fig3}, bottom-right).

%% Figure 4 %%
\begin{figure*}
 \noindent\includegraphics[width=39pc,angle=0]{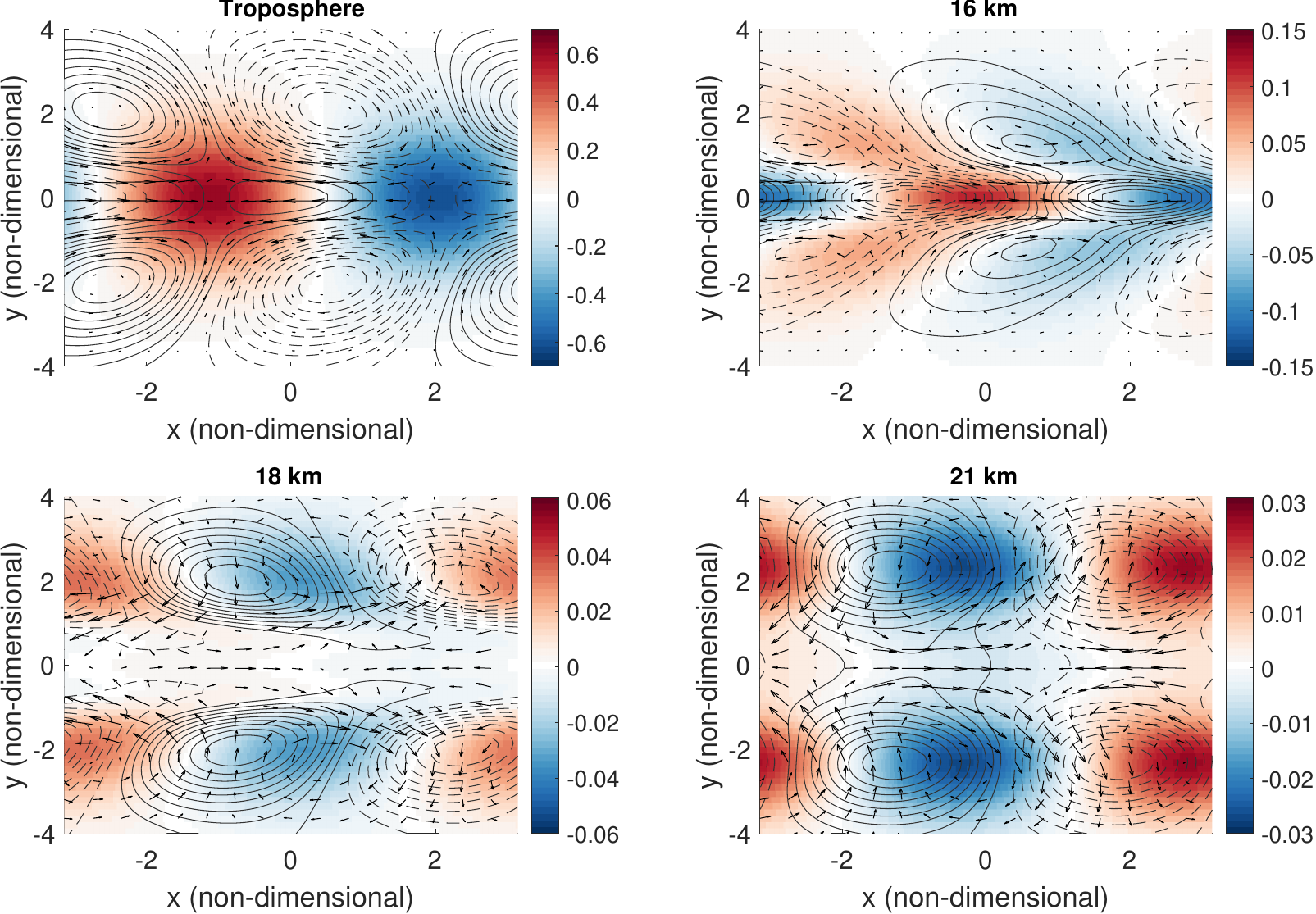}
 \caption{Similar to Figure \ref{fig3} but for the case of westerly mean wind ($U = 1$) in the stratosphere.} \label{fig4}
\end{figure*}

On the other hand, Figure \ref{fig4} shows horizontal cross-sections for the MJO-like mode, but now for mean westerly winds in the stratosphere, $U = 1$. While the boundary-layer eigenmode is unchanged, the stratospheric wave patterns are significantly different from the case with mean stratospheric easterlies. Now, the equatorial Kelvin-wave component of the solution is strongly damped in the stratosphere, and a clear signature of an upward propagating equatorial Rossby wave is evident.

These solutions unsurprisingly show that the stratospheric wind can play a prominent role in modulating tropospheric upwards wave propagation. We can investigate this quantitatively, by calculating the total stratospheric wave energy flux, defined as $\overline{\phi^\prime w^\prime}$ integrated over the entire stratospheric numerical domain, where averaging occurs zonally. This quantity is positive for all eigenmodes in this study, as wave energy must be propagating upwards. Under mean easterly flow, the Kelvin wave component has increased vertical energy flux, while the opposite is true for the Rossby wave component (not shown). Conversely, the Kelvin wave component has strongly damped vertical energy flux under mean westerly flow, while that of the Rossby wave is significantly increased (not shown). These results are consistent with those predicted by linear theory of equatorial Rossby wave propagation under mean easterly flow [see equation 4.7.21 in \citet{andrews1987middle}], and linear theory of Kelvin wave propagation under mean westerly flow [the Doppler-shifted phase speed of the Kelvin-wave must be eastward, as in equation 4.7.10 in \citet{andrews1987middle}].

% \begin{equation}
%     \int_{-\infty}^\infty \int_1^\infty \int_{-\pi}^\pi \phi^\prime w^\prime \: dx \: dz^* \: dy
% \end{equation}

%% Figure 5 %%
\begin{figure*}
 \noindent\includegraphics[width=39pc,angle=0]{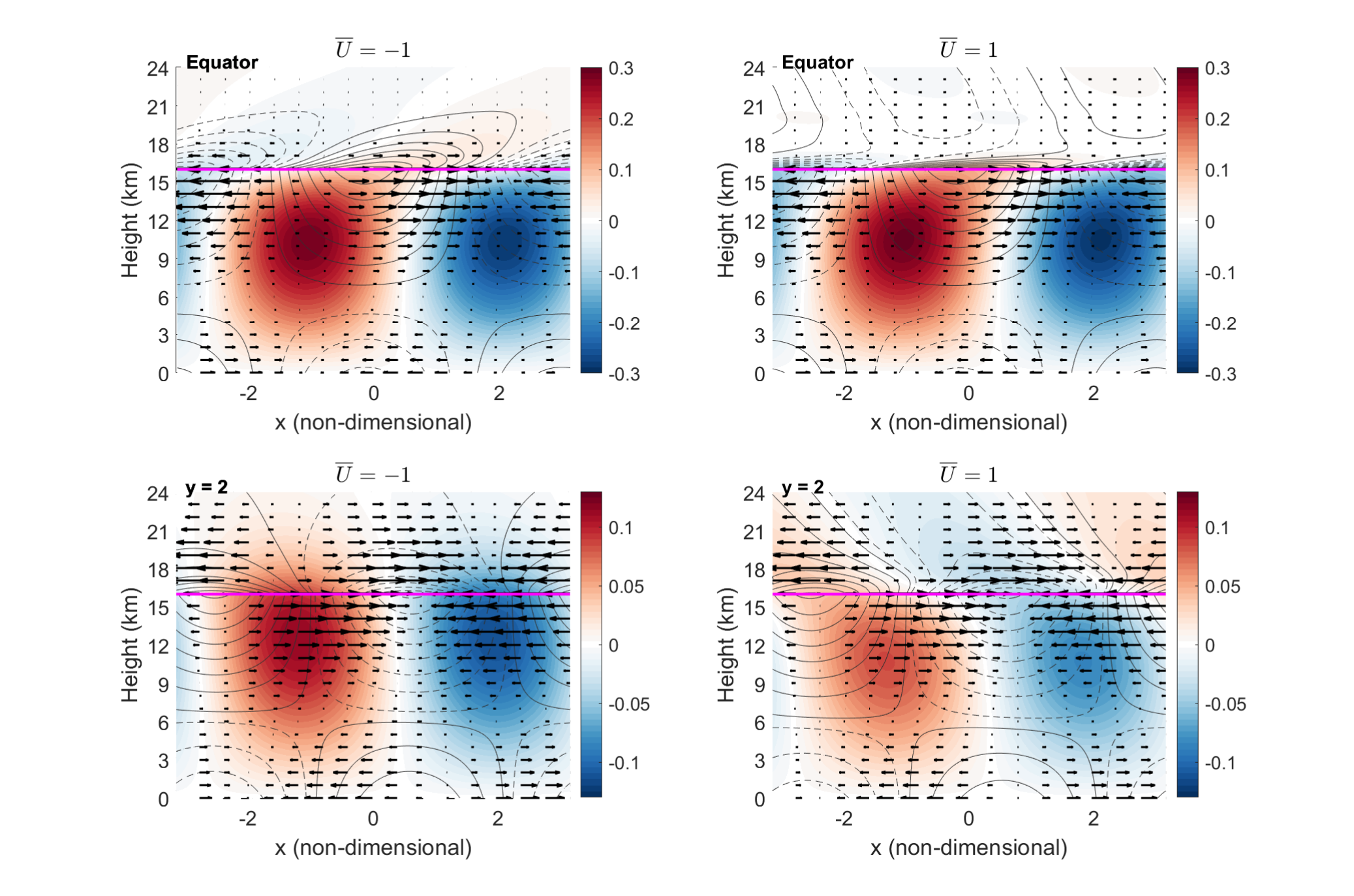}
 \caption{Vertical-zonal cross-sections of the $k = 1$, MJO-like eigenmode, for the case of capped, constant shear in stratospheric wind, under (left column) easterlies [$U = -1$] and (right column) westerlies [$U = 1$]. Cross-sections are at meridional locations of the (top) equator, and (bottom) $y = 2$ (around 20$^\circ$ latitude). As in Figure \ref{fig3}, the tropopause is set to 16-km, indicated by the magenta line. Contours indicate the geopotential perturbations, where solid (dashed) lines indicate positive (negative) perturbations. Arrows indicate wind perturbations, and color shadings indicate vertical velocity perturbations (positive for upward) at the level indicated in the label. Non-dimensional parameters selected are $\alpha = 0.35$, $\chi = 0.2$, $\gamma = 5$, $D = 0.25$, $G = 0.02$, $\delta_x = 30$, $S = 40$, $C = 1.25$, $C_i = 0$.} \label{fig5}
\end{figure*}

Vertical-zonal cross sections of the MJO-like mode also show strong dependence of the vertical structure on the stratospheric wind. Figure \ref{fig5} shows a vertical cross section at the equator and at $y = 2$ (around 20$^\circ$ latitude), for the MJO-like mode under both stratospheric easterlies and westerlies. On the equator, as shown in Figure \ref{fig5}, top row, the tropospheric vertical structure projects heavily onto the first baroclinic mode, since a pure first barolinic mode structure has $w$ maximizing at $\approx 10$ km. Furthermore, an eastward tilt with height exists on the equator, due to the presence of the upward propagating Kelvin wave. However, under mean westerly winds, the upward propagation of the equatorial Kelvin-wave is strongly damped, as evidenced in Figure \ref{fig5}, top-right, in comparison to that under the easterly case (c.f. Figure \ref{fig5}, top-left). The explanation for the varying phase tilts can be understood through the linear dynamics. In a linear model, the phase speed of the MJO-like mode in the troposphere must be equal to the Doppler-shifted phase speed of the stratospheric wave:
\begin{equation}
     c_{\text{mjo}} = \overline{U}_s + c_{\text{wave}}
\end{equation}
Suppose that $c_{\text{mjo}}$ is more or less fixed by tropospheric dynamics. Then, under mean easterly flow, $c_{\text{wave}}$ must increase to match $c_{\text{mjo}}$. A larger phase speed is associated with a larger vertical wavelength. Observational data indeed suggests that there is an upward propagating Kelvin wave associated with the MJO in the stratosphere, and that the QBO modulates the strength and propagation characteristics of this Kelvin wave \citep{nishimoto2017influence}.

The vertical-zonal cross sections at $y = 2$ show a much different pattern, since at these latitudes, the Kelvin wave signal is much weaker than the Rossby wave signal. At $y = 2$, the vertical tilt with height in the stratosphere becomes westward, indicating the presence of upward propagating Rossby waves. This may explain the stratospheric westward tilt in height observed by \citet{hendon2018differences} in their composites of the MJO during westerly phases of the QBO. The westward phase tilt is stronger with stratospheric westerlies than easterlies, indicating Doppler-shifting of the stratospheric Rossby waves. Interestingly, the tropospheric vertical structure at $y = 2$ is also dependent on the stratospheric zonal wind. Most notably, the barotropic mode is much stronger at $y = 2$ than on the equator, indicated by the vertical velocity peaking at $z \approx 13$ km rather than at $z \approx 10$ km. The Rossby gyres becoming increasingly dominated by the barotropic mode as one moves poleward; in fact at $y = 3$, the Rossby gyres are almost completely barotropic (not shown). As explained by \citet{lin2021effect}, the magnitude of the baroclinic mode decays more quickly with distance from the equator than that of the barotropic mode, leading to barotropic Rossby vortices in the subtropics and extratropics. Furthermore, at $y = 2$, the magnitude of the barotropic mode (and associated vertical velocities) is stronger under stratospheric easterlies than westerlies. Dopper-shifting of the Rossby gyres is clear: under stratospheric easterlies, the westward phase tilt with height decreases as the phase speed of the Rossby wave decreases, while under westerlies, the phase tilt with height must increase as the phase speed of the Rossby wave increases. As such, vertical velocities in the upper troposphere are stronger under stratospheric easterlies than westerlies. The presence of these barotropic Rossby gyres has been found in three-dimensional observational composites of the MJO \citep{adames2015three}. One important caveat, however, is that  Equation (\ref{eq_us_linear}), while simple, assumes that the stratospheric zonal wind shown has no meridional dependence. The QBO does not have a large meridional extent, certainly not extending to the location of the Rossby gyres, which is an issue that will be remedied in the next section \citep{baldwin2001quasi}.

Despite significant stratospheric control on upward wave propagation, the growth rate, frequency, and total upwards energy flux for the MJO-like mode are nearly constant across the different stratospheric wind profiles (not shown). In general, the MJO-like mode has a slow phase speed ($\approx 5$~m~s$^{-1}$) and a small total upward energy flux, such that any changes in upwards wave propagation have negligible effects on the overall wave characteristics. Regardless, these results show that the MJO-like eigenmode in the linear model resembles the observed MJO, and, at-least in the linear framework, is able to excite modes that resemble equatorial Kelvin and Rossby waves in the stratosphere. While the sign of the stratospheric wind has a minimal effect on growth rate and frequency, it strongly influences the upper tropospheric and lower stratospheric wave patterns, which could ultimately influence the behavior of cirrus clouds. Since cirrus clouds can significantly modulate tropospheric radiative heating, stratospheric influence on cirrus clouds deserves further exploration.

\subsection{Ice cloud radiative forcing of the MJO}
% In this section, show that the cirrus cloud parameterization reproduces the old parameterization
Since the MJO's upper-tropospheric wave pattern can be strongly dependent on the sign of stratospheric wind, it is reasonable to think that the stratosphere influences TTL cirrus clouds. To investigate this hypothesis, we will use the simple representation of cirrus clouds and their effect on radiative cooling, shown in Equations (\ref{eq_qv}) and (\ref{eq_q}), under zero stratospheric mean wind, $\overline{U}_s = 0$.

%% Figure 6 %%
\begin{figure*}
 \noindent\includegraphics[width=39pc,angle=0]{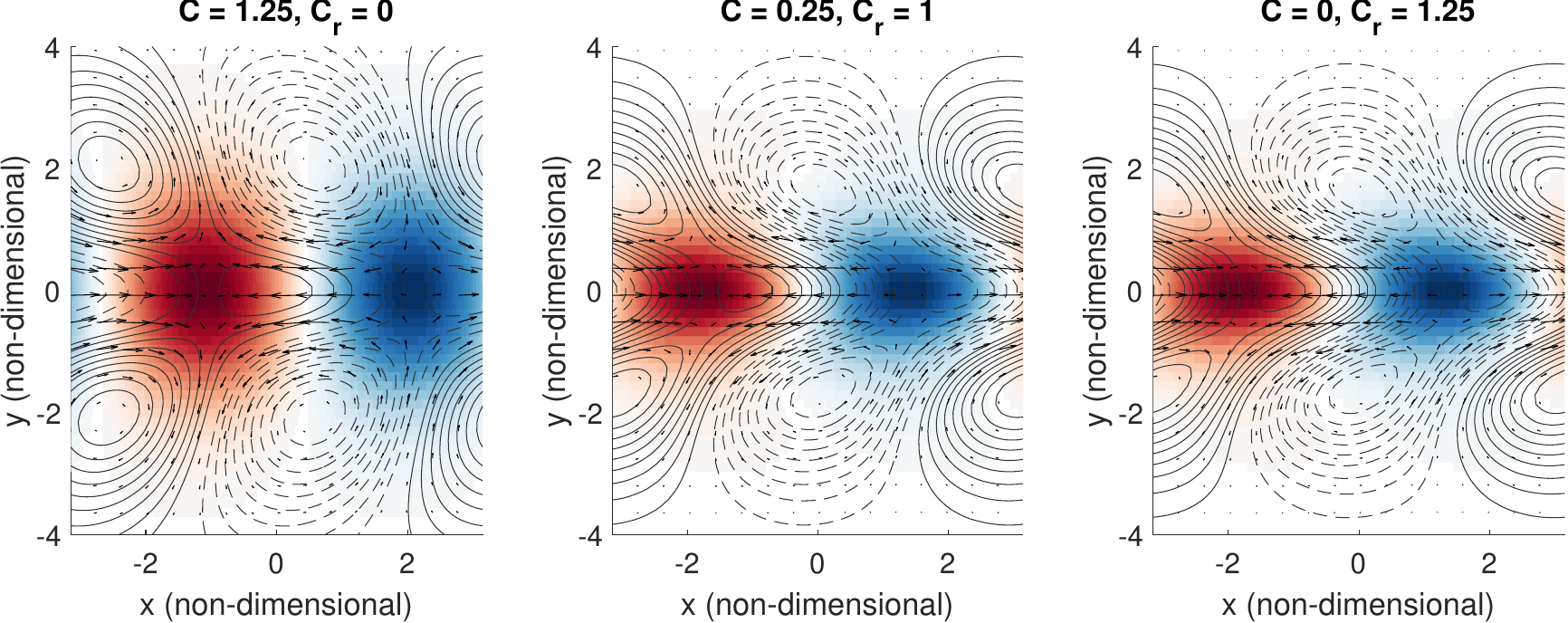}
 \caption{Horizontal cross-sections of the $k = 1$, MJO-like eigenmode at the boundary layer, under (left) the original cloud radiative feedback parameterization of \citet{khairoutdinov2018intraseasonal}, (middle) a parameterization where the ice clouds radiative feedback is much larger than that of water vapor, and (right) one where only ice clouds contribute to radiative cooling perturbations. $z_c = 14$ km for where $C_i > 0$, and $U_s = 0$. Contours indicate the saturation entropy in the troposphere, where solid (dashed) lines indicate positive (negative) perturbations. Arrows indicate wind perturbations, and color shadings indicate mid-level vertical velocity perturbations (positive for upward). Non-dimensional parameters selected are $\alpha = 0.35$, $\chi = 0.2$, $\gamma = 5$, $D = 0.25$, $G = 0.02$, $\delta_x = 30$, $S = 40$.} \label{fig6}
\end{figure*}

First, it would be prudent to show that the cirrus cloud parameterization produces eigenmodes that have very similar horizontal structures to the MJO-like eigenmodes that appear under the original cloud-radiative feedback parameterization of \citet{khairoutdinov2018intraseasonal}. Figure \ref{fig6} shows the horizontal summary eigenfunction of the $k = 1$, eastward propagating, MJO-like mode, under a zero-mean zonal wind in the stratosphere, but for varying magnitudes of $C$ and $C_i$. We first select $z_c = 14$ km, which is the vertical level at which the climatological cloud fraction peaks. Sensitivity to $z_c$ will be discussed later in this section. The eigenmode for the ``realistic" case of $C = 0.25$ and $C_i = 1$, where ice clouds are responsible for most of the cloud-radiative feedback, as informed from the radiative transfer calculations, is shown in Figure \ref{fig6}, center. The horizontal structure is qualitatively similar to that of the eigenmode using the original cloud-radiative feedback equation of \citet{khairoutdinov2018intraseasonal}, except for the Rossby gyres being located slightly more equatorwards. In addition, there are not significant differences in the horizontal structure when completely replacing the cloud-radiative feedback equation from \citet{khairoutdinov2018intraseasonal} with the ice-cloud parameterization (Figure \ref{fig6}, right).

%% Figure 7 %%
\begin{figure*}
 \noindent\includegraphics[width=39pc,angle=0]{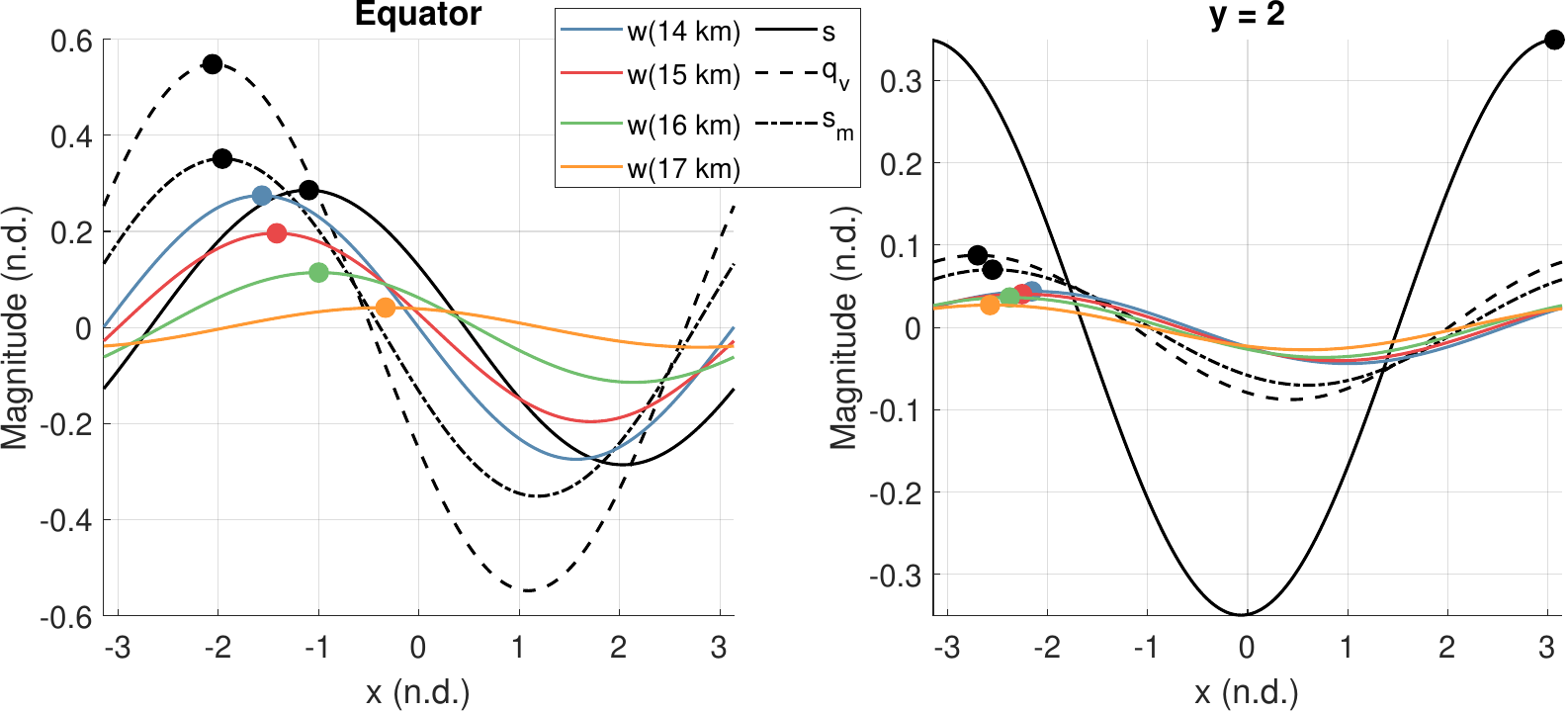}
 \caption{(Left) Wave patterns along the equator of $s$, $s_m$, $q_v$, and $w$ at various vertical levels, for the $k = 1$, MJO-like mode, with $C = 0.25$, $C_i = 1$, $z_c = 14$ km, and $U_s = 0$. Solid dots denote where the wave pattern reaches its maximum value. The tropopause height is $H = 16$ km, and the stratospheric zonal wind is zero. Non-dimensional parameters selected are $\alpha = 0.35$, $\chi = 0.2$, $\gamma = 5$, $D = 0.25$, $G = 0.02$, $\delta_x = 30$, $S = 40$. (Right) Same as left but at $y = 2$.} \label{fig7}
\end{figure*}

The reason why the ice-cloud parameterization does not significantly change the MJO structure is straightforward to understand. Figure \ref{fig7} shows the wave pattern along the equator and $y = 2$, under the case of zero-mean stratospheric wind, using $z_c = 14$ km. In both zonal cross-sections, the phase of $q_v$ is nearly coincident with the phase of $s_m$, such that modifying the relative magnitudes of $C$ and $C_i$ will only lead to small changes in the phase relationship between total radiative cooling and saturation entropy (solid line). This is expected, as $s_m$ in the tropics mostly represents a moisture deficit. The near collocation of low-level water vapor and upper tropospheric cirrus clouds in the linear model is consistent with the same observation in MJO composites, as shown in Figure \ref{fig1}. On the equator, we see a eastward tilt with height in vertical velocity in the upper troposphere and lower stratosphere, a consequence of the upward propagation of the equatorial Kelvin wave. On the other hand, at $y = 2$, there is a nearly barotropic structure in vertical velocity, as explained earlier.

%% Figure 8 %%
\begin{figure}[t]
 \center \noindent \includegraphics[width=19pc,angle=0]{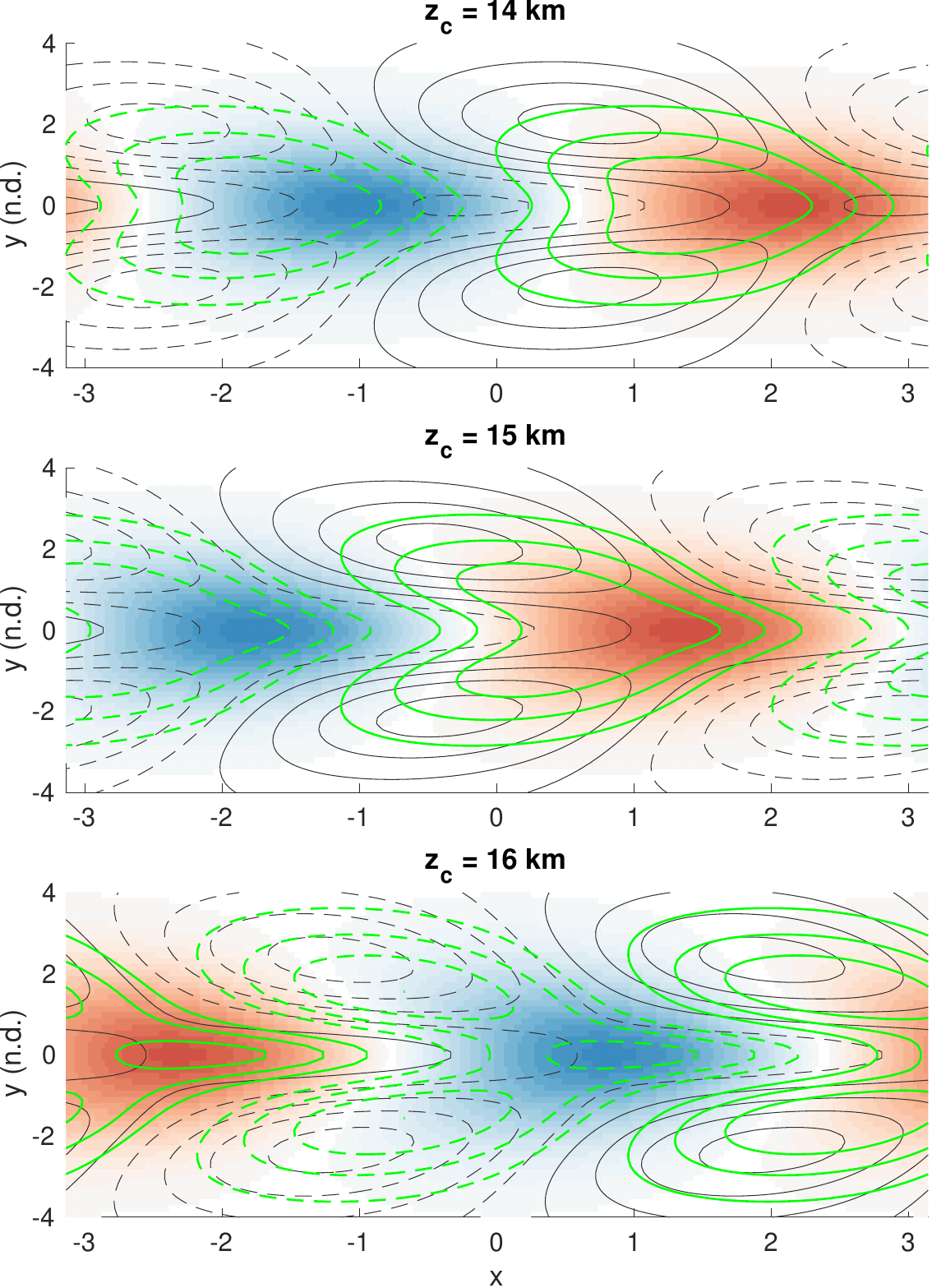}
 \caption{Horizontal cross-sections of the $k = 1$, MJO-like eigenmode at the boundary layer, using (top) $z_c = 14$ km, (center) $z_c = 15$ km, (bottom) $z_c = 16$ km, with zero stratospheric mean wind. Non-dimensional cloud-radiative feedback parameters selected are $C = 0.25$ and $C_i = 1$. Black contours indicate the saturation entropy, while green contours indicate the cirrus cloud cover ($q_v$), where solid (dashed) lines indicate positive (negative) perturbations. Color shadings indicate mid-level vertical velocity perturbations (positive for upward), with a range of [-1, 1].} \label{fig8}
\end{figure}

Here, it is worth commenting on the cirrus level $z_c$, as well as the horizontal structure of $q_v$. Although we know $z_c$ should be confined to the upper troposphere, the behavior of the MJO-like mode under varying stratospheric winds can also be modulated by the level at which we set $z_c$. It is not clear what value of $z_c$ is most realistic, and as such, we shall present the behavior of the solutions using a range of $z_c$ throughout this study. In order to understand the effect of varying $z_c$ without the influence of the stratospheric zonal wind, we look at the MJO-like eigenmode structure under $z_c = 14$, $15$, and $16$ km. We choose the non-dimensional parameters of $C = 0.25$ and $C_i = 1$, where ice-clouds dominate the cloud radiative feedback, as informed by the radiative transfer calculations. Figure \ref{fig8} shows that as $z_c$ shifts higher in the troposphere, the equatorial Kelvin-wave like component of the eigenmode shifts eastwards with respect to the Rossby gyres, which themselves become more prominent. Regardless of the value of $z_c$, convection maximizes on the equator, though off-equator convection associated with the Rossby gyres becomes stronger as $z_c$ increases. For $z_c = 14$ km, cirrus cloud cover is largely centered on the equator, with a Kelvin-wave like structure. As $z_c$ increases, the poleward extent of cirrus cloud cover increases, extending further into the off-equator Rossby gyres. These properties are reasonable given the differences in vertical structure between the Rossby and Kelvin waves. Since the Rossby gyres are nearly barotropic, vertical velocity increases with height, such that increasing $z_c$ increases off-equator cirrus cloud fraction, further destabilizing the Rossby gyres. On the other hand, the equatorial Kelvin wave component of the MJO-like mode is primarily baroclinic. Thus, vertical velocity and cirrus cloud radiative forcing decrease with increasing $z_c$. Furthermore, vertical velocity has an eastward tilt with height on the equator, such that as $z_c$ increases in height, the cirrus cloud radiative forcing shifts eastward on the equator. Therefore, as $z_c$ increases in height, the cirrus cloud radiative forcing shifts eastward on the equator and is enhanced off the equator, leading to a ``phase decoupling" between the Rossby and Kelvin wave. Note that MJO-composites of TTL cirrus show that during the MJO, cirrus clouds fraction anomalies maximize on the equator, and extend poleward around 25$^\circ$ latitude, with a Rossby-gyre like signal off the equator \citep{virts2014observations}. This makes the cirrus cloud pattern for $z_c = 15$ km and $z = 16$ km more comparable to our observations, which is around the height where cirrus cloud anomalies associated with the MJO peak, at least in the analysis performed by \citet{del2015cloud}.
% While changing $z_c$ does modify the horizontal structure of the MJO-like eigenmode,
% a modified equatorial Rossby wave \citep{ahmed2021mjo}.
% As alluded to earlier, an vertically integrated metric involving $q_v$ is perhaps more directly related to column radiative heating perturbations.

\subsubsection{Dynamical modulation of cirrus clouds}
Now, we can use the linear model to understand how stratospheric modulation of the MJO's upward wave propagation (and the associated vertical velocity anomalies) can influence cirrus cloud formation. In the linear model, this is done through the term on the RHS of Equation (\ref{eq_qv}), which represents the dynamical contribution of upward propagating waves to cirrus cloud formation. Before imposing stratospheric zonal wind, we must graduate to a realistic, QBO-like oscillation in mean zonal wind, which will vary the sign and limit the meridional extent of mean-zonal wind in the stratosphere [see Appendix for details]. We run a set of experiments varying both $z_c$ and the sign and magnitude of the QBO wind pattern [Equation (\ref{eq_qbo_wind})], but do not include any advection of cirrus clouds ($U_c = 0$). Again, we choose $C = 0.25$ and $C_i = 1$.

%% Figure 9 %%
\begin{figure}[t]
 \center \noindent\includegraphics[width=19pc,angle=0]{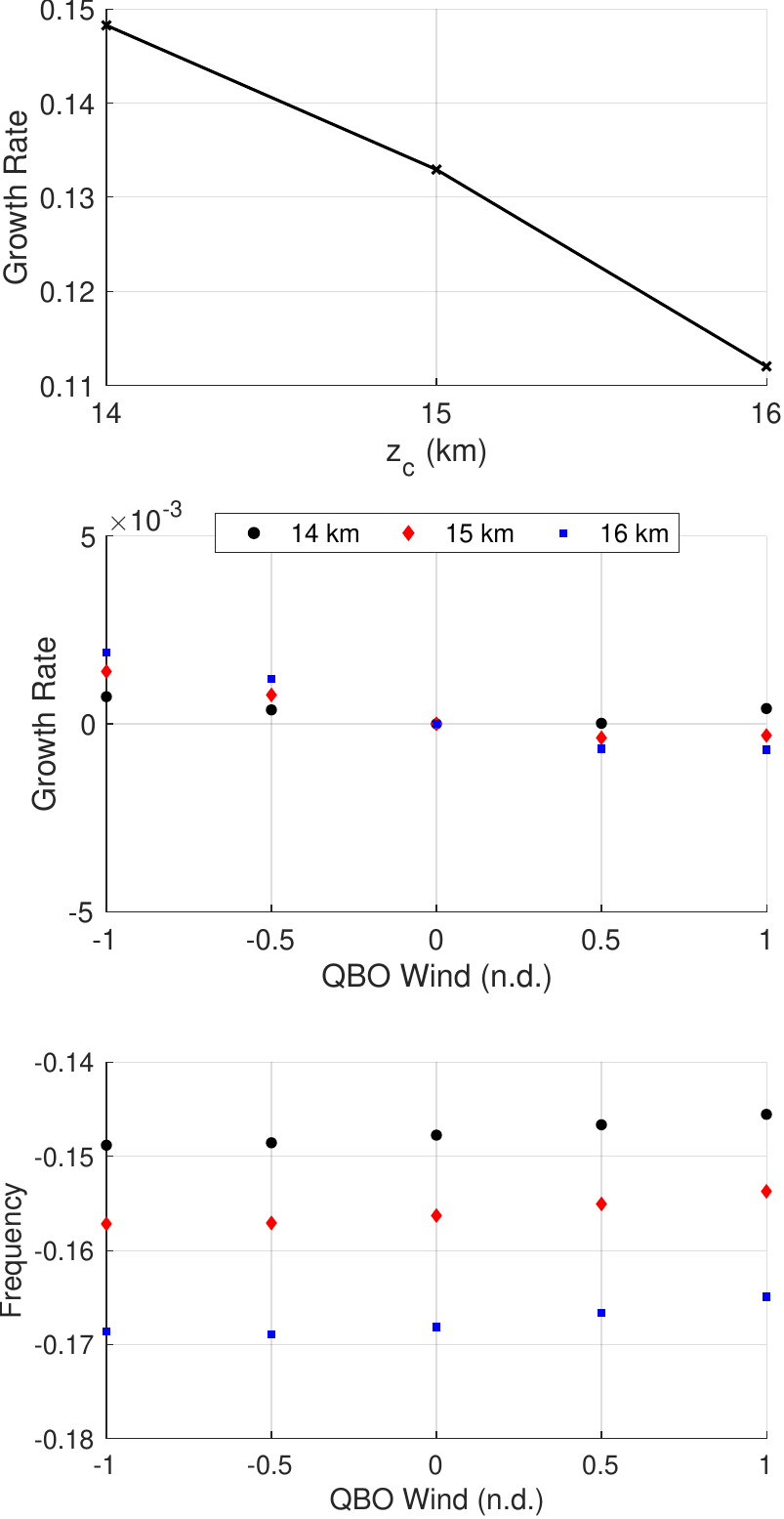}
 \caption{(Top) Growth rate of the $k = 1$, MJO-like eigenmode under varying $z_c$, $C = 0.25$, $C_i = 1$, and zero-mean stratospheric wind. (Middle) Difference in growth rate from that of zero-mean stratospheric wind [see top panel], under varying QBO phases and $z_c$. (Bottom) Frequencies of the MJO-like mode under varying QBO phases and $z_c$. All growth rates and frequencies are non-dimensional.} \label{fig9}
\end{figure}

Figure \ref{fig9} shows the growth rate and frequency of the MJO-like mode under varying stratospheric winds, and $z_c = 14$, $15$, and $16$ km. The growth rates under zero-mean stratospheric wind (Figure \ref{fig9}, top panel) decrease with increasing $z_c$, since $\Upsilon$ is fixed across the experiments and $w$ generally decreases with height in the troposphere. However, the absolute magnitude of the growth rates are of less importance, since the growth rates under zero-mean stratospheric wind can be easily modified by adjusting $\Upsilon$ and $C_i$, both of which have large uncertainties because of cloud macrophysical and microphysical processes.

Rather, experiments in which the stratospheric wind is varied can inform us on how the phase of the QBO modulates the linear growth rate. Figure \ref{fig9}, middle panel, shows that the growth rates are higher for easterly phases of the QBO than westerly phases, but only when $z_c$ is higher than 14 km, which is consistent with the observation that the MJO is stronger during QBO easterlies than westerlies \citep{yoo2016modulation}. On the other hand, the growth rates are nearly constant with QBO phase for $z_c = 14$ km. The phase speeds of the MJO-like mode are slightly faster under QBO easterlies than westerlies for all choices of $z_c$. Note that this is inconsistent with observations, which seem to indicate that the MJO propagates faster under QBOW than QBOE \citep{nishimoto2017influence}, though, as noted by \citet{son2017stratospheric}, stronger MJO events propagate more slowly across the Maritime Continent than weaker ones.

%% Figure 10 %%
\begin{figure*}
 \noindent\includegraphics[width=39pc,angle=0]{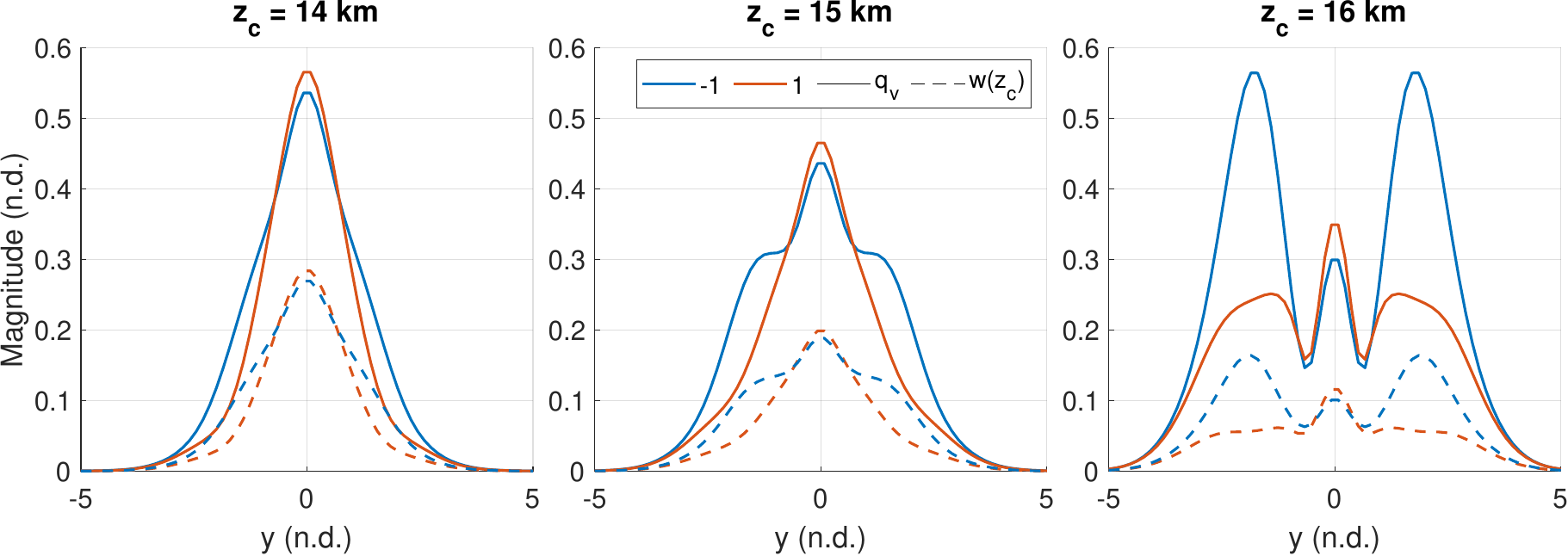}
 \caption{Meridional dependence of anomalies of (solid) $|q_v|$ and (dashed) $|w(z_c)|$ for the $k = 1$, MJO-like eigenmode under varying $z_c$ and QBO wind. Blue line indicates easterly QBO wind, and red line indicates westerly QBO wind.} \label{fig10}
\end{figure*}

How can we explain the results of the linear model? To understand these differences, we look at $q_v$ and $w(z_c)$ under varying $z_c$ and stratospheric wind (Figure \ref{fig10}), since these variables control the cloud-radiative feedback in the system. At $z_c = 14$ km, there are small differences in the magnitude of $w(z_c)$ between QBO easterlies and westerlies. Hence, there are minimal differences in $q_v$ and the cloud-radiative feedback, leading to minimal control of the growth rate by the stratospheric wind. On the other hand, as $z_c$ increases, we see a poleward expansion in $w(z_c)$; at $z_c = 16$ km, there are three local peaks in $w(z_c)$ and $q_v$, coinciding with the two off-equatorial Rossby gyres and the equatorial Kelvin-wave.

The behavior of $w(z_c)$ with increasing $z_c$ is directly attributable to the differences in vertical structure between the equatorial Kelvin-wave and Rossby wave components of the MJO-like eigenmode. Since the Rossby waves are much more barotropic under lower stratospheric easterlies than westerlies, off-equator $w(z_c)$ and thus $q_v$ are generally larger under QBOE than QBOW. This effect is stronger with larger $z_c$ given the barotropic vertical structure of the Rossby waves. On the other hand, the first baroclinic mode, which is less sensitive to the mean zonal wind of the stratosphere, dominates the vertical structure of the Kelvin-wave. Thus, even though the Kelvin-wave is trapped in the troposphere under stratospheric westerlies, resulting in a larger vertical velocity on the equator for westerly stratospheric winds, the difference in $w(z_c)$ is not significant given the dominance of the baroclinic mode over the barotropic mode near the equator.

%Most importantly, however, the stratosphere wind also modifies the upward propagation of Kelvin and Rossby waves. Under westerly stratospheric winds, Kelvin waves are tropospherically trapped, while under easterly stratospheric winds, Rossby waves are tropospherically trapped. The differences are stark when comparing $w(z_c)$ at a fixed $z_c$ for varying stratospheric wind. For instance, for $z_c = 16$ km, the off-equator ($y \approx 2$) vertical velocity is larger under stratospheric easterlies than westerlies, because the Rossby wave propagates upwards less efficiently. The same effect can be seen in the equatorial vertical velocities, albeit at a lesser effect; the Kelvin-wave is trapped in the troposphere under stratospheric westerlies, resulting in a larger vertical velocity on the equator for westerly stratospheric winds.

Stratospheric influence on our cirrus cloud proxy and the growth rate of the MJO-like eigenmode can be understood quantitatively by using the entropy variance equation, which can be obtained by multiplying Equation (\ref{eq_s}) by $s$ and averaging over the domain:
\begin{equation}
    \pder[]{t} \{ s^2 \} = (1 + C) \{s s_m \} + C_i \{q_v s\} - \{ w s \} - \alpha \{ u_b s \} - \chi \{ s^2 \} \label{eq_s_energy} \\
\end{equation}
where $\{ \}$ is an averaging operator defined in \citet{emanuel2020slow} as:
\begin{equation}
    \{ f \} = \int_{-\infty}^{\infty} \int_0^{2\pi} f dx dy
\end{equation}

%% Figure 11 %%
\begin{figure}[t]
 \center \noindent\includegraphics[width=19pc,angle=0]{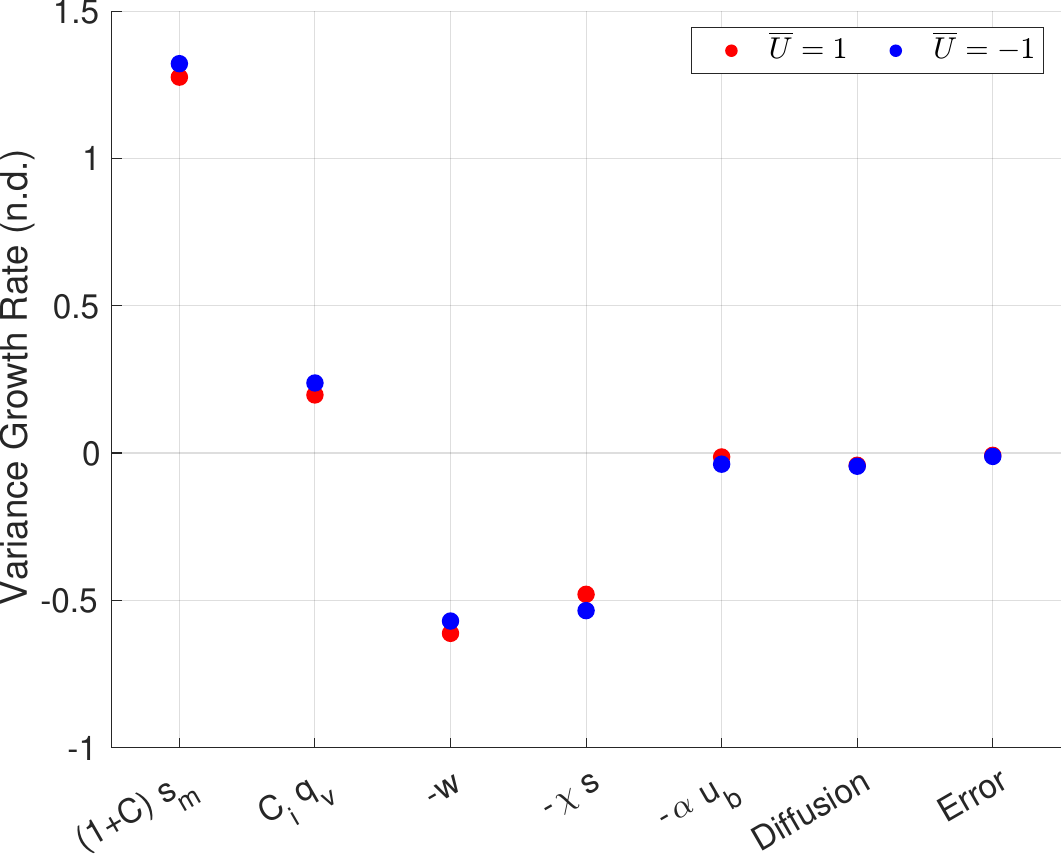}
 \caption{Decomposition of the time-dependent energy budget for saturation entropy, as outlined in Equation (\ref{eq_s_energy}), for the MJO-like eigenmode under (blue) QBOE and (red) QBOE, using $z_c = 15$ km. Term labels indicate the domain-average correlations between the term and $s$. For comparison, eigenmodes are normalized to have the same total domain energy. Diffusion represents weak meridional diffusion required for numerical stability, and error represents numerical error.} \label{fig11}
\end{figure}

In Equation (\ref{eq_s_energy}), the first two terms on the right hand side represent cloud-radiative feedbacks, and are positive if clouds destabilize the MJO-like mode. The third term is damping from large-scale vertical motion, and is negative since $s$ and $w$ are positively correlated. The fourth term represents the WISHE feedback, while the fifth term is an entropy damping term that is negative definite \citep{emanuel2020slow}. Numerical diffusion and error are not explicitly included in the budget, but are shown in ensuing figures for completeness. Figure \ref{fig11} shows the contribution of each term in Equation (\ref{eq_s_energy}), under QBO easterlies and westerlies. The eigenmodes under stratospheric easterlies and westerlies are normalized to have the same total domain energy prior to computation of the energy budget, in order for comparisons to be meaningful. The energy budget shows that cloud-radiative feedbacks are highly destabilizing for the MJO-like mode, with stronger feedbacks under QBO easterlies than westerlies. WISHE, on the other hand, plays a small role in energy growth, though it plays a crucial role in eastward propagation \citep{khairoutdinov2018intraseasonal, emanuel2020slow}.

These results suggest that the stratosphere can play a crucial role in modulating the tropospheric vertical structure of the MJO, most notably through modulation of vertical wave propagation. Stratospheric modulation of tropospheric vertical structure can be significant, since small changes to the large-scale vertical velocity can have large effects on cirrus clouds in the TTL. Understanding how the behavior of cirrus clouds are modified by the stratosphere seems to be crucial to understanding the MJO-QBO relationship, since perturbations to tropospheric radiative cooling are dominated by cirrus clouds.

% An important question to ask is if this mechanism generalizes to other equatorial waves? As shown by \citep{abhik2019sensitivity}, only the MJO, and perhaps the convectively coupled Kelvin wave, is modulated by the QBO. The convectively coupled Kelvin wave has a much weaker cloud-radiative feedback than the MJO, which help could explain this contradiction \citep{sakaeda2020unique}.

% TODO need to cite ltierature on detrainment level of cirrus clouds from convection, as opposed to TTL cirrus clouds... Sassen et al. (2009).
% Hartman and Larson (2002): Fixed anvil hypothesis
% Figure 7 in Virts and Wallace (2014): cloud fraction signature associated with MJO's Rossby gyres

\subsubsection{Influence of stratospheric stratification}
In general, the QBO's direct modulation of the dry stratification in the lower stratosphere has been the most studied pathway through which the QBO can influence the MJO \citep{son2017stratospheric, abhik2019sensitivity, sakaeda2020unique}. In order to get insight into this suggested pathway, we use the linear model to explore how changes to the mean-state dry stratification in the stratosphere ($S$) influence the growth rates of the MJO-like mode. In this linear model, a smaller stratospheric $N_2$ generally increases the ``leakiness" of the MJO-like mode. In other words, more wave-energy is lost to the stratosphere, which tends to dampen the growth rate of waves [see \citet{lin2021effect} for more details]. However, when $N_2$ decreases, the magnitude of the barotropic mode also increases, which tends to increase the magnitude of the large-scale vertical velocity in the upper troposphere (increasing the ``top-heaviness" of $w$), which can increase the ice cloud fraction through dynamical forcing. These two effects can counter-act each other, and it is not obvious which effect wins out in the end.

%% Figure 12 %%
\begin{figure*}
 \noindent\includegraphics[width=39pc,angle=0]{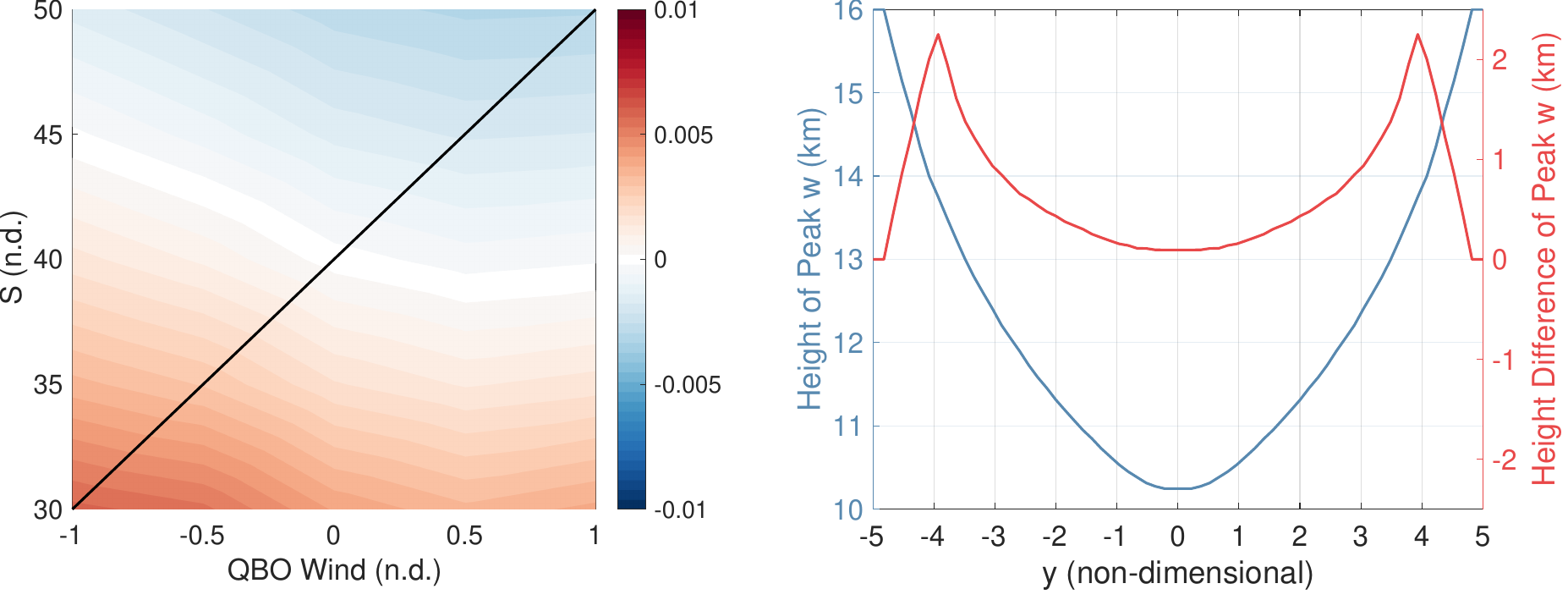}
 \caption{(Left) Dependence of growth rate of the MJO-like mode on stratospheric dry stratification ($S$) and stratospheric wind. Growth rates are shown with respect to anomalies from the growth rate with zero-mean stratospheric wind, and $S = 40$. Black line shows the approximate relationship between the QBO phase/magnitude and the dry stratification in the lower stratosphere. (Right) Height of peak vertical velocity for case of $S = 30$ and zero-mean stratospheric wind, and the difference in the height of peak vertical velocity between the $S=30$ and $S=50$ cases, under zero-mean stratospheric wind.} \label{fig12}
\end{figure*}

In the stratosphere, the QBO modulates the dry stratification in the lower stratosphere by around 10\% of it's mean value \citep{nishimoto2017influence}. Using our linear model, we perturbed $S$ with similar orders of magnitude, and investigated the dependence of the growth rates on both $S$ and the lower stratospheric wind. Quite interestingly, we see in Figure \ref{fig12} that decreasing the stratosphere stratification actually leads to larger growth rates of the MJO-like mode. In particular, the vertical velocity profile becomes more ``top-heavy", as shown in Figure \ref{fig12}, right. Therefore, while more wave-energy is lost to the stratosphere with decreasing $S$ (not shown), the magnitude of the barotropic mode increases, leading to an upward shift in the vertical velocity profile and to increased cloud-radiative forcing from cirrus clouds (not shown). This upward shift of the vertical velocity profile during QBOE has some support from observations \citep{sakaeda2020unique}, though the exact mechanism through which this occurs is not quite clear yet. Note, however, that the way in which these effects occur in this linear model is through a modulation of the barotropic mode magnitude (large-scale dynamics), not through a modulation of convective instability, the latter of which is the predominant thinking of the QBO-MJO pathway. Regardless, these results from the linear model are interesting in their own right, and at-least support the idea that QBO modulation of the lower-stratospheric dry stratification can influence the MJO magnitude.

\subsubsection{Zonal advection by the background wind}
As mentioned earlier, an eastward tilt with height in cirrus cloud fraction (Figure \ref{fig1}) may also be the result of zonal advection by the background wind. Anomalous zonal advection may play a role in determining the phase relationship of radiative heating anomalies with saturation entropy anomalies. The bulk background zonal advection of cirrus clouds is the subject of attention in this section, and quantified by the second term on the LHS of Equation (\ref{eq_qv}). Note, that in the original formulation of \citet{khairoutdinov2018intraseasonal}, the mean wind of the troposphere is set to zero, such that advective processes are ignored, the largest approximation of which is likely the omission of the horizontal advection of moisture. While the focus of this study is upper-tropospheric/lower-stratospheric dynamics, horizontal advection of moisture can be significant, especially in light of more recent moisture-mode based theories of MJO-propagation, which highlight the importance of horizontal advection of moisture \citep{ahmed2021mjo, wang2022unified}. While we include horizontal advection of cirrus clouds, this is not inconsistent with the simultaneous omission of advective processes of column integrated moist entropy. This is because ice clouds occur in the upper troposphere, where the absolute magnitude of perturbation mixing ratios are small, such that ice clouds do not make a substantial contribution to the column integrated entropy, as opposed to water vapor perturbations in the mid-levels.

%% Figure 13 %%
\begin{figure*}
 \noindent\includegraphics[width=39pc,angle=0]{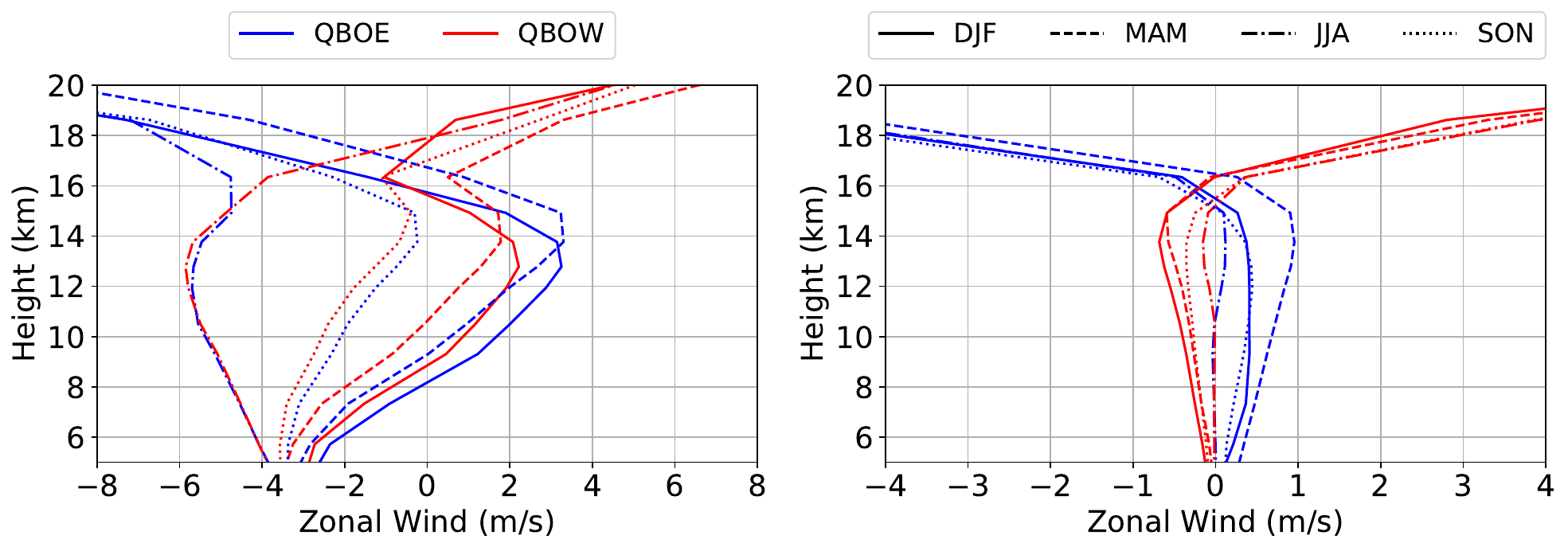}
 \caption{(Left) Tropical averaged ($10^{\circ}S-10^{\circ}N$) zonal wind, separated into (blue) easterly and (red) westerly phases of the QBO, as well as (solid) DJF, (dashed) MAM, (dot-dashed) JJA, and (dotted) SON. (Right) Same as left but for deseasonalized, zonal wind anomalies. Zonal winds are calculated using 1979-2020 ERA5 re-analysis fields.} \label{fig13}
\end{figure*}

To understand how to represent $U_c$, we turn to re-analysis data. Figure \ref{fig13} shows the tropical averaged (10$^\circ$S-10$^\circ$N), mean and anomalous zonal wind, separated into different seasons and easterly and westerly phases of the QBO. During boreal winter, the season where the MJO amplitude is strongest and the MJO-QBO relationship is observed, there are upper tropospheric mean westerlies in the tropics, regardless of the QBO phase. The presence of upper tropospheric westerlies may advect cirrus clouds associated with MJO-convection eastward, leading to an eastward tilt with height. However, the strength of the tropical-averaged TTL westerlies is slightly weaker (around 1-1.5~m~s$^{-1}$) during QBOW than QBOE. This is most evident in the deseasonalized zonal wind anomalies shown in Figure \ref{fig13}, right. During lower stratospheric westerlies (QBOW), easterly anomalies exist in the upper troposphere, while the opposite is true during lower stratospheric easterlies (QBOE). Differential advection of upper tropospheric cirrus clouds between QBO phases may play a role in modulating the strength of the MJO. It is important to note that while zonal-means of the zonal wind have been taken since we are linearizing about the tropical basic state, the upper-tropospheric zonal wind does vary significantly zonally, and is predominantly easterly in the Indo-Pacific warmpool. Thus, these results have regional dependency; the magnitude of the anomalous upper-tropospheric zonal winds that are opposite signed of the QBO phase is smaller from 0$^\circ$E-180$^\circ$E than from 180$^\circ$E-360$^\circ$E (not shown), the former of which encompasses the warm pool region.

While it may be hard to believe that differences of 1~m~s$^{-1}$ can make large differences in MJO strength, the anomalies are not negligible with respect to the magnitude of the mean winds (which themselves are tropical averages). However, it is important to note that there are still upper tropospheric westerlies during MAM, and that the difference in the magnitude of the westerlies between QBOE and QBOW is larger than that during DJF ($\approx 1.5$~m~s$^{-1}$). This is at odds with the fact that the MJO-QBO relationship is only observed during boreal winter \citep{yoo2016modulation}, though the seasonality of the strength of the MJO (the MJO being strongest in boreal winter) may also play a role \citep{zhang2004seasonality}. During JJA and SON, there are pronounced upper tropospheric easterlies in the tropics; much of the easterly signal in the tropics is due to the presence of the upper tropospheric anticyclone associated with the South Asian monsoon.

The zonal wind profile in boreal winter leads us to include anomalous zonal advection that is opposite signed of the QBO, albeit at a much smaller magnitude. We define the non-dimensional $U_c$ as follows:
\begin{equation}
    U_c = 0.1 - U_s / 20 \label{eq_u_adv}
\end{equation}
This definition of $U_c$ has mean westerlies, as in boreal winter, and is also consistent with the fact that upper-tropospheric zonal wind anomalies are opposite signed from the phase of the QBO occur, and approximately 20 times smaller than the typical amplitude of the QBO. Note that $U_c$ is positive within the range of $U_s$ used in this study.

%% Figure 14 %%
\begin{figure}[t]
 \center \noindent\includegraphics[width=19pc,angle=0]{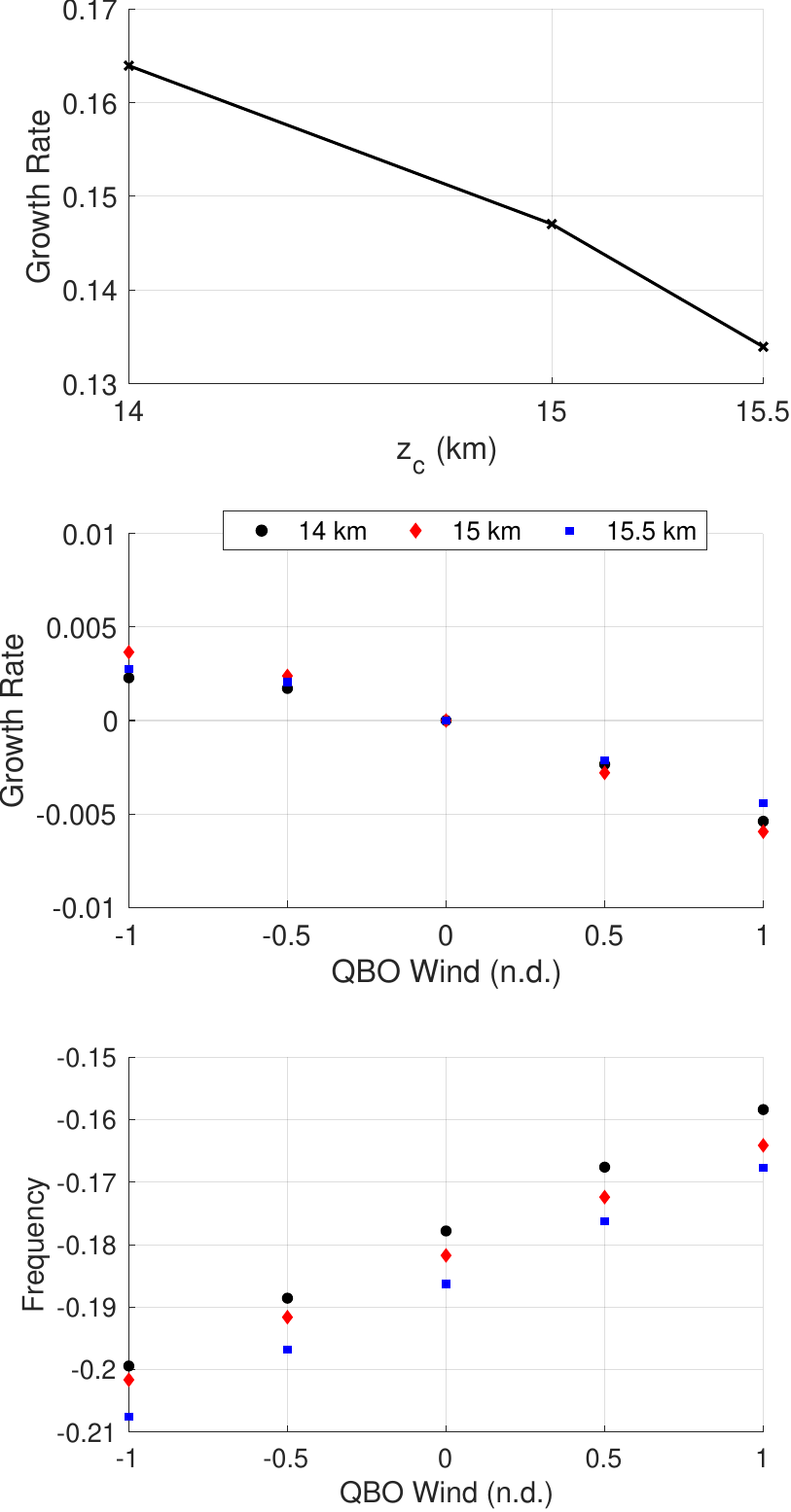}
 \caption{Same as Figure \ref{fig9} but now with zonal advection as defined in the text.} \label{fig14}
\end{figure}

\begin{figure*}
 \noindent\includegraphics[width=39pc,angle=0]{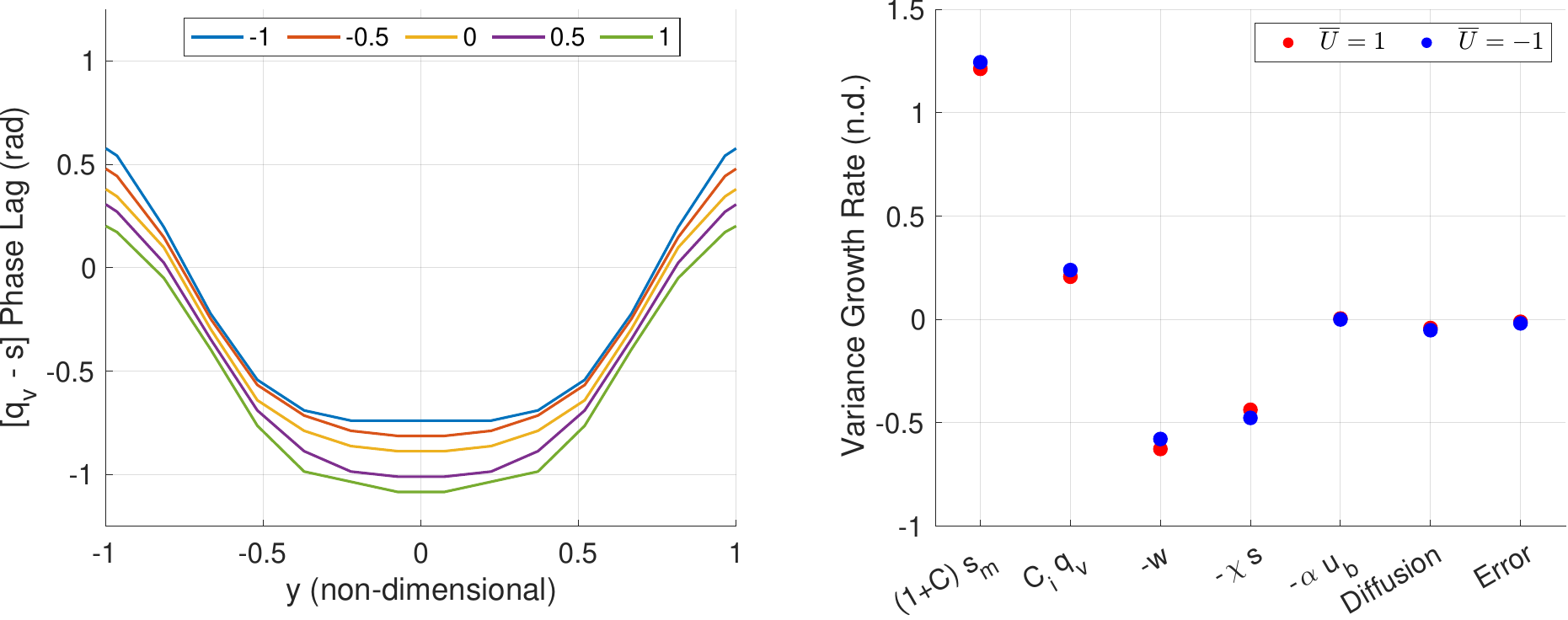}
 \caption{(Left) Meridional dependence of the phase lag, in radians, between cirrus clouds ($q_v$) and saturation entropy, for the $k = 1$, MJO-like eigenmode, using $z_c = 15$ km, under varying stratospheric QBO wind ($U$) and anomalous zonal advection of cirrus clouds following Equation (\ref{eq_u_adv}). (Right) Same as Figure \ref{fig11} but now for the case including anomalous zonal advection of cirrus clouds.} \label{fig15}
\end{figure*}

When including zonal advection of cirrus clouds, the tropospheric eigenmode of the MJO-like mode generally retains the familiar shape of an equatorial Kelvin wave lagged and flanked by a Rossby wave (not shown). Figure \ref{fig14} shows the growth rate and frequency of the $k = 1$ MJO-like mode, but now with the inclusion of weak zonal advection of cirrus clouds, according to Equation (\ref{eq_u_adv}).  We see that growth rates are higher with upper tropospheric westerly advection (easterly QBO) as compared to easterly advection (westerly QBO). The differences in growth rates increase with the strength of advection, and the magnitude of these differences are largely the same across the range of $z_c$. Note that here we use $z_c = 15.5$ km in lieu of of $z_c = 16$ km, since the tropopause is assumed to have a zero-mean zonal wind. These results can be explained when looking at the relationship between saturation entropy and $q_v$, which forces the system through radiative heating perturbations. In the linear model, zonal advection acts primarily to shift the phase relationship between cirrus clouds and saturation entropy. This is seen in Figure \ref{fig15}, where the phase lag between $s$ and $q_v$ near the equator is reduced under QBO easterlies by anomalous eastward advection. On the other hand, $s$ and $q_v$ are more out of phase when subjected to anomalous westward advection. As a result, the mode grows faster and propagates faster under anomalous westward cirrus cloud advection. The opposite is true of anomalous eastward advection. This is seen clearly in the saturation energy budget shown in Figure \ref{fig15} (right), where destabilization from cloud-radiation feedbacks is stronger under anomalous eastward advection (QBOE) than westward advection (QBOW).

%% Figure 15 %%

These results show that even very weak zonal advection of cirrus clouds can significantly modulate the MJO. After all, cirrus clouds dominate the cloud-radiative forcing for the MJO. As such, it seems crucial to understand what processes are responsible for the eastward tilt with height of cirrus cloud fraction on the equator, as shown in Figure \ref{fig1}. If the eastward tilt arises from mean westerly advection by the background wind, QBO-induced anomalous zonal advection could be a pathway through which the QBO modulates the MJO. While the linear model formulated in this paper highlights this potential pathway, it can only crudely represent cloud processes. More research, especially numerical modeling, is necessary to validate the hypotheses outlined in this study.

\section{Summary and discussion  \label{sec_summary}}
This study aims to better understand the effect of the stratosphere on the MJO, primarily motivated by observations of the modulation of the MJO by the QBO \citep{yoo2016modulation}. First, we created composites of ice cloud fraction and mid-level water vapor during aggregated MJO phases, showing that anomalies of ice cloud fraction and lower-tropospheric water vapor are nearly-collocated with each other. These composites also show that near the equator, there is an eastward tilt with height in cloud fraction with respect to the horizontal maximum in lower-tropospheric water vapor, which is hypothesized to be driven by the upward propagating Kelvin wave, and/or mean westerly advection by the background flow. These composites also show that OLR is strongly correlated with both lower-tropospheric water vapor and ice-cloud fraction. In order to understand the order of magnitude effect of each quantity on OLR, we used vertical profiles of anomalous cloud fraction and water vapor from the MJO composites as input into RRTM. Radiative transfer modeling showed that ice-clouds in the upper troposphere dominate the radiative forcing, as measured by OLR.

Given the importance of cirrus clouds on radiative heating perturbations, as well as the potential for the behavior of cirrus clouds to be modulated by the stratosphere, we incorporated a simple prognostic equation for cirrus clouds into the coupled troposphere-stratosphere linear model described in \citet{lin2021effect}. In our linear model, cirrus clouds are forced dynamically and allowed to be advected by the background zonal wind. The representation of cirrus clouds occurs at a single level, $z_c$, which is modified throughout the study. In order to investigate stratospheric influence on the MJO, the model in \citet{lin2021effect} is further extended to include a non-zero stratospheric mean wind in thermal wind balance.

We use the mean-state over the TOGA-COARE IOP to inform the non-dimensional parameters of the linear model, which show an MJO-like growing mode. We focus on how this MJO-like mode in the linear system interacts with the stratosphere. Specifically, we analyze stratosphere-induced perturbations to cirrus cloud fraction and the ensuing modification to radiative heating perturbations. The behavior of the MJO-like mode was analyzed under two zonal wind profiles in the stratosphere, from constant shear to a more realistic, QBO-like oscillation in zonal wind. As in \citet{lin2021effect}, a numerical model is used to solve for growth rate and phase speed of the MJO-like mode by integrating the equations forward in time. The main findings of the study are summarized below:
\begin{itemize}
    \item Eastward tilts with height in MJO-associated ice cloud fraction are observed above $\approx 14$~km, near the equator, from CALIPSO cloud occurrence profiles. The eastward tilt with height can be explained through dynamical forcing via the upwards propagating Kelvin wave, eastward advection by the mean zonal flow in the upper troposphere, or both.
    \item Stratospheric wind can also play a large role in modifying the tropospheric vertical structure of the MJO-like mode, primarily through changes to upward wave propagation. Under stratospheric westerlies, the upward propagation of the equatorial Kelvin wave associated with the MJO is strongly damped, as predicted by linear theory. MJO-associated westward (eastward) tilts with height under westerlies (easterlies), as shown by \citet{hendon2018differences}, can be explained by upward propagating Rossby (Kelvin) waves associated with the MJO. In the absence of cloud-ice processes, the QBO phase has an insignificant effect on MJO growth rate and frequency.
    \item A variety of tropospheric vertical structures can be explained by the superposition of the barotropic and first baroclinic modes. In this linear model, the MJO-like mode's equatorial Kelvin wave component is dominated by the first baroclinic mode whereas the off-equatorial Rossby wave component is strongly barotropic. These results agree with three-dimensional observational composites of the MJO \citep{adames2015three}.
    \item A simple representation of cirrus-clouds and their associated feedback on tropospheric radiative cooling are incorporated into the linear model. When cirrus clouds are allowed to be forced dynamically, we obtain a MJO-like eigenmode as a linear solution. The eigenmode is shown to be similar to the MJO-like eigenmode under the original cloud radiation feedback parameterization of \citet{khairoutdinov2018intraseasonal}, in which radiative cooling is assumed to be related to the mid-level moisture deficit. The horizontal structure of ice clouds in the MJO-like eigenmode is similar to MJO-composites of upper-tropospheric cirrus clouds \citep{virts2014observations}.
    \item When cirrus clouds are dynamical forced, growth rates of the MJO-like eigenmode are shown to be stronger under stratospheric easterlies, as compared to stratospheric westerlies, consistent with observational evidence that the MJO is stronger under QBOE. This behavior is attributed to dynamical modulation of cirrus clouds by upward propagating waves; the Rossby gyres have a stronger barotropic mode under stratospheric easterlies, which enhances cirrus cloud forcing in the upper troposphere and strengthens MJO destabilization by radiative heating perturbations.
    \item Tropical-averaged upper-tropospheric zonal winds are shown to be mean westerly during boreal winter, but anomalously westerly (easterly) under QBOE (QBOW).  The influence of anomalous advection of cirrus clouds by the background flow is investigated by including zonal advection in the cirrus-cloud prognostic equation. We show that QBOE-associated anomalous westerly zonal advection in the upper troposphere also enhances the growth rate of the MJO-like mode, by shifting the phase of radiative heating to be more in phase with saturation entropy anomalies.
\end{itemize}

% TODO: Talk about seasonality of MJO relationship, and why the relationship only occurs post 1979
There are certain aspects of the MJO-QBO relationship that were not thoroughly explored in this study but deserve attention. As briefly discussed, the MJO-QBO relationship only appears during boreal winter \citep{yoo2016modulation}. During boreal winter, the MJO is much closer to and symmetric about the equator than during boreal summer, where slowly propagating intraseasonal variability takes the form of northwest-southeast oriented, northward propagating bands called the Boreal Summer Intraseasonal Oscillation (BSISO) \citep{adames2016seasonality, kikuchi2021boreal}. Since the QBO is confined in its meridional extent, it is possible that the seasonality of the MJO-QBO relationship is closely tied to the displacement of BSISO off the equator. This connection was not explored in this study, though could be investigated using unified models of the MJO and BSISO \citep{wang2022unified}. Furthermore, the MJO-QBO relationship is only significant after 1979 \citep{sakaeda2020unique}. If cirrus clouds and their modulation of tropospheric radiation on intraseasonal time scales indeed play a role in the MJO-QBO relationship, then reliable satellite observations of OLR would be necessary in order to capture the MJO-QBO connection in re-analysis \citep{liebmann1996description}. Of course, this is conjecture, and far from conclusive.

While this study focused on the $k = 1$ MJO-like mode, we also investigated the aforementioned mechanisms in the $k = 2$ (and higher) MJO-like modes, and the results are worth mentioning here. In general, the MJO-like modes propagate more slowly as the horizontal wavenumber increases, and hence the magnitude of both the barotropic mode and wave energy loss to the stratosphere decreases with zonal wavenumber \citep{lin2021effect}. This means that the differences in growth rates between stratospheric easterlies or westerlies are diminished when only considering changes to the vertical energy flux. We also performed experiments looking at differences in growth rates from modulation of the cirrus-cloud feedback. Dynamical modulation of cirrus is reduced for the smaller scale MJO-like modes, since the dynamical forcing ($w$) is smaller in magnitude than the dynamical forcing for the $k = 1$ mode. On the other hand, modulation of the growth rates of the MJO-like mode through zonal advection in the upper troposphere is still significant for the smaller scale MJO-like modes.

The linear model formulated in this study serves as a step towards better understanding tropospheric-stratospheric coupling in the tropics. One may rightfully question the extent to which linear models can capture the true relationship between the MJO and QBO. Non-linear wave dynamics and wave-breaking at critical layers, which our linear model fails to resolve, might be important components of the MJO-QBO relationship. After all, the QBO owes its existence to momentum transfer to the mean flow from breaking upward propagating waves \citep{lindzen1968theory}. There is also some evidence that upward propagating waves in the lower stratosphere often become disconnected from the space-time forcing of the troposphere, which would invalidate assumptions of linearity \citep{yang2012influence}.

In this study, we also assume that there is a discontinuous transition between a convecting troposphere and a passive stratospheric at a specified surface. This idealization may affect the results shown in this study, since in reality, the TTL serves as the interface between these two dynamical regimes \citep{fueglistaler2009tropical}. The presence of the TTL may change the behavior of the barotropic mode and vertical tilt in the MJO-like mode, through changes to the index of refraction. Thus, focus on the exact value of $z_c$ may not be as important, since the behavior of this model could change if a TTL were included in this linear model.

Throughout this study, we have also shown that stratospheric influence on the growth rate of the MJO-like mode can depend quite strongly on $z_c$ (as well as other non-dimensional cloud parameters such as $\Upsilon$). This highlights the importance that cloud physics has in the MJO-QBO connection, and may point to why modeling efforts have so far failed to capture this connection; a small spread in the general characteristics of cirrus clouds in climate models can produce a large spread in models' abilities to capture the MJO-QBO relationship. Yet, one may also hesitate at our simple parameterization of cirrus clouds, which only considers a single level $z_c$ to be of importance for cirrus cloud radiative feedbacks. In reality, the net radiative forcing by high-clouds is a complex, non-linear function of optical depth and cloud-top height \citep{fu2002tropical}. As such, radiative heating perturbations are better represented using depth-integrated quantities, which we neglected in the spirit of simplicity, though this will be the subject of future work. The results in this study may be sensitive to this behavior. Our cirrus cloud parameterization also reduces cloud microphysical and macrophysical processes to a couple linear relations. Thus, it is worth commenting on the sensitivity of the results to the magnitude of $\Upsilon$, which controls water vapor production, and that of $C_i$, which controls cirrus cloud radiative forcing. While $C_i$ is empirically constrained in this study, the value of $\Upsilon$ is selected fairly arbitrarily. In general, increasing $\Upsilon$ increases the magnitude of $q_v$ anomalies and subsequently the cloud radiative forcing, increasing the difference in the MJO-like mode's growth rates under stratospheric westerlies and easterlies. This behavior is as expected, as it places greater weight on cirrus cloud radiative forcing. Regardless, future work will focus on rigorous validation of the cirrus cloud parameterization, which will also include constraining $\Upsilon$. If cirrus cloud modulation is important, it is unlikely to be captured by GCMs, owing to coarse vertical resolution and possibly to microphysics parameterizations.

When modeling complex phenomena in the atmosphere, it is often necessary to make simplifying assumptions to make tractable progress on understanding the underlying dynamics. Thus, the results of this theoretical study should be viewed through a lens of skepticism. But, the interpretations could prove to be a useful guide for high-resolution modeling experiments. This will be the subject of future work.

%%%%%%%%%%%%%%%%%%%%%%%%%%%%%%%%%%%%%%%%%%%%%%%%%%%%%%%%%%%%%%%%%%%%%
% TABLES---INSERT NEAR IN-TEXT DISCUSSION
%%%%%%%%%%%%%%%%%%%%%%%%%%%%%%%%%%%%%%%%%%%%%%%%%%%%%%%%%%%%%%%%%%%%%
%%  Enter tables near where they are discussed within the document.
%%  Please place tables before/after paragraphs, not within a paragraph.
%%
%
%\begin{table}[t]
%\caption{This is a sample table caption and table layout.  Enter as many tables as
%  necessary at the end of your manuscript. Table from Lorenz (1963).}\label{t1}
%\begin{center}
%\begin{tabular}{ccccrrcrc}
%\hline\hline
%$N$ & $X$ & $Y$ & $Z$\\
%\hline
% 0000 & 0000 & 0010 & 0000 \\
% 0005 & 0004 & 0012 & 0000 \\
% 0010 & 0009 & 0020 & 0000 \\
% 0015 & 0016 & 0036 & 0002 \\
% 0020 & 0030 & 0066 & 0007 \\
% 0025 & 0054 & 0115 & 0024 \\
%\hline
%\end{tabular}
%\end{center}
%\end{table}

%%%%%%%%%%%%%%%%%%%%%%%%%%%%%%%%%%%%%%%%%%%%%%%%%%%%%%%%%%%%%%%%%%%%%
% FIGURES---INSERT NEAR IN-TEXT DISCUSSION
%%%%%%%%%%%%%%%%%%%%%%%%%%%%%%%%%%%%%%%%%%%%%%%%%%%%%%%%%%%%%%%%%%%%%
%%  Enter figures near where they are discussed within the document.
%%  Please place figures before/after paragraphs, not within a paragraph.
% %
%
%\begin{figure}[t]
%  \noindent\includegraphics[width=19pc,angle=0]{figure01.pdf}\\
%  \caption{Enter the caption for your figure here.  Repeat as
%  necessary for each of your figures. Figure from \protect\cite{Knutti2008}.}\label{f1}
%\end{figure}

\clearpage
%%%%%%%%%%%%%%%%%%%%%%%%%%%%%%%%%%%%%%%%%%%%%%%%%%%%%%%%%%%%%%%%%%%%%
% ACKNOWLEDGMENTS
%%%%%%%%%%%%%%%%%%%%%%%%%%%%%%%%%%%%%%%%%%%%%%%%%%%%%%%%%%%%%%%%%%%%%
\acknowledgments The authors gratefully acknowledge the support of the National Science Foundation through grant NSF ICER-1854929. The first author also thanks Tristan Abbott for providing python bindings for RRTMG.

%  Keep acknowledgments (note correct spelling: no ``e'' between the ``g'' and
% ``m'') as brief as possible. In general, acknowledge only direct help in
%  writing or research. Financial support (e.g., grant numbers) for the work done,
%  for an author, or for the laboratory where the work was performed must be
%  acknowledged here rather than as footnotes to the title or to an author's name.
%  Contribution numbers (if the work has been published by the author's institution
%  or organization) should be placed in the acknowledgments rather than as
%  footnotes to the title or to an author's name.

%%%%%%%%%%%%%%%%%%%%%%%%%%%%%%%%%%%%%%%%%%%%%%%%%%%%%%%%%%%%%%%%%%%%%
% DATA AVAILABILITY STATEMENT
%%%%%%%%%%%%%%%%%%%%%%%%%%%%%%%%%%%%%%%%%%%%%%%%%%%%%%%%%%%%%%%%%%%%%
%
%
\datastatement ERA5 re-analysis is publicly available through the Climate Data Store of the Copernicus Climate Change Service. The MJO OMI index is available online at \url{https://psl.noaa.gov/mjo/mjoindex/omi.1x.txt}. NOAA's Interpolated OLR dataset is available at \url{https://psl.noaa.gov/data/gridded/data.interp_OLR.html}. The CALIOP data is available at \url{https://asdc.larc.nasa.gov/project/CALIPSO/CAL_LID_L3_Cloud_Occurrence-Standard-V1-00_V1-00}. Model source code, instructions to run the model, and code to generate the figures from outputs of the numerical model are available at github.com/linjonathan. All code and figures were generated using Python and MatLab.
%  The data availability statement is where authors should describe how the data underlying
%  the findings within the article can be accessed and reused. Authors should attempt to
%  provide unrestricted access to all data and materials underlying reported findings.
%  If data access is restricted, authors must mention this in the statement. See
%  {http://www.ametsoc.org/PubsDataPolicy} for more info.

%%%%%%%%%%%%%%%%%%%%%%%%%%%%%%%%%%%%%%%%%%%%%%%%%%%%%%%%%%%%%%%%%%%%%
% APPENDIXES
%%%%%%%%%%%%%%%%%%%%%%%%%%%%%%%%%%%%%%%%%%%%%%%%%%%%%%%%%%%%%%%%%%%%%
%
%% If only one appendix, use
\appendix

%% If more than one appendix, use \appendix[<letter>], e.g.,
%\appendix[A]

%% Appendix title is necessary! For appendix title:
\appendixtitle{Details of Model Formulation}
\subsection*{Non-dimensionalization}
Here, we define the non-dimensional scalings for the relevant variables that appear in the linear model. The scalings for the tropospheric quantities are identical to those described in the appendix of \citet{khairoutdinov2018intraseasonal}, and the scalings for the stratospheric quantities are identical to those described in the appendix of \citet{lin2021effect}.
\begin{align}
    t &\rightarrow \frac{a}{\beta L_y^2} \\
    x &\rightarrow a x \\
    w^\prime &\rightarrow C_k |\textbf{V}| w \\
    q^\prime_v &\rightarrow \overline{q_i^*} q_v \\
    \overline{U}_s &\rightarrow \beta L_y^2 U_s \\
    T &\rightarrow \frac{\beta^2 L_y^4}{R_d} T
\end{align}
where the terms on the left of the arrow are the dimensional quantities, and those on the right are the non-dimensional quantities. With these non-dimensionalizations, we have:
\begin{align}
    \Gamma_m = \frac{C_k |\overline{V}| a}{\beta L_y^2 H}
\end{align}
% \begin{equation}
%     \dot{Q}^\prime = C \frac{\overline{s_0} - \overline{s_{b}}}{(\overline{s_{b}} - \overline{s_m})^2} \epsilon \pder[\overline{s_d}]{z} C_k |\overline{\textbf{V}}| s_m^\prime
% \end{equation}

\subsection*{QBO Formulation}

\begin{figure*}
 \center
 \noindent\includegraphics[width=39pc,angle=0]{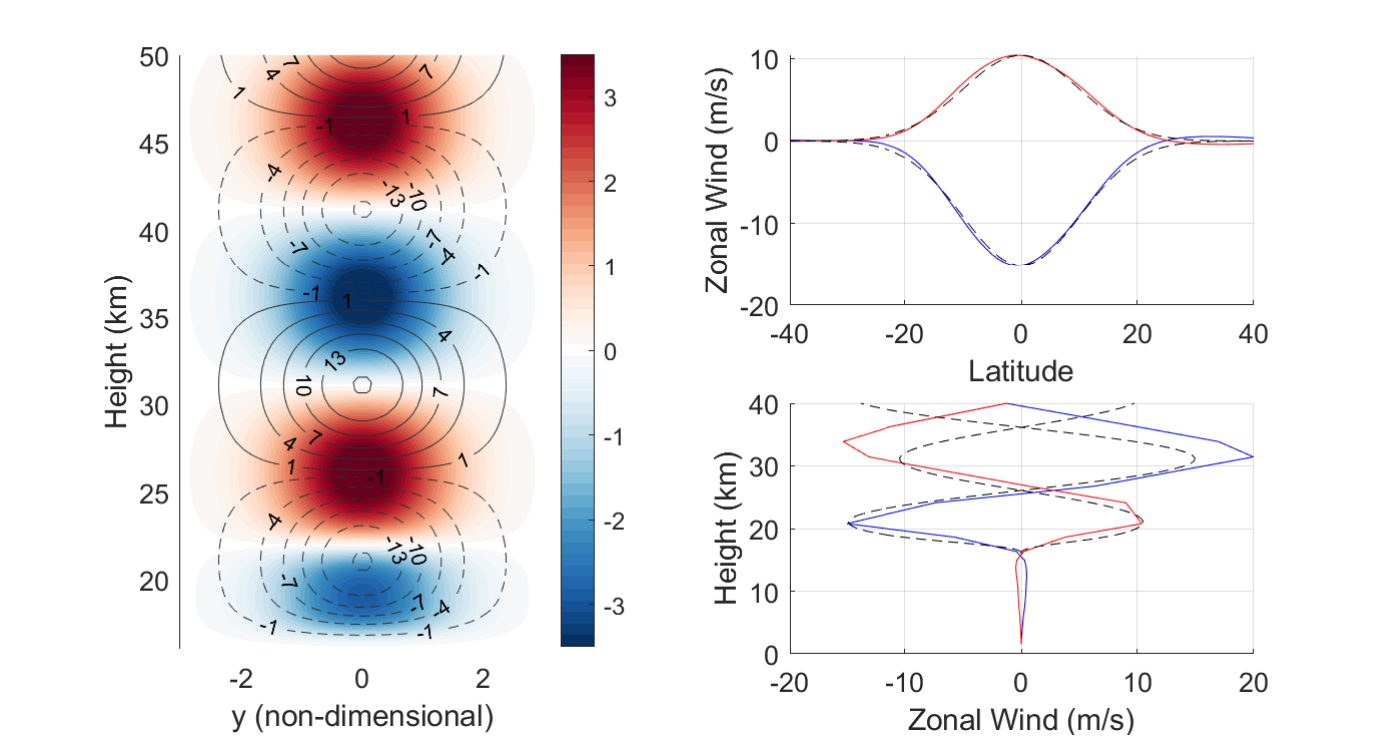}
 \caption{(Left) Example of imposed QBO-easterly mean-state in the stratosphere, with contours indicating dimensional zonal wind speed ($m$~$s^{-1}$), and shading indicating associated dimensional temperature anomalies ($K$). Contour intervals are in spacing of $3$~$m$~$s^{-1}$, starting at $\pm 1$~$m$~$s^{-1}$. Non-dimensional parameters are $U = -0.5$, $b_1 = 5$, $b_2 = 0.5$, $b_3 = 0.01$, and the tropopause is set to 16 km. (Top right) Meridional dependence of the zonally averaged, anomalous zonal wind during (blue) QBOE and (red) QBOW phases from ERA5 re-analysis 1979-2020, with dashed black lines indicating the dimensional, meridional dependence of the zonal wind in the linear model, arbitrarily scaled for zonal wind magnitude. (Bottom right) Same as top right but for the vertical structure of the anomalous zonal wind during (blue) QBOE and (red) QBOW, with height.} \label{fa1}
\end{figure*}

The mathematical form of the mean wind we impose is:
\begin{equation}
    U_s(y, z^*) = U \: R \:  \sin \big( b_1 (z^* - 1) \big) \exp \big( -b_2 y^2 \big) \label{eq_qbo_wind}
\end{equation}
where $b_1$, $b_2$, and $b_3$ are non-dimensional constants that control the vertical wavelength of the oscillation, meridional extent of the mean wind, and vertical extent of the damping factor, respectively. $U$ is the maximum magnitude of the mean wind, and $R$ is a non-dimensional damping factor that is only active in the lower stratosphere and ensures that there is no temperature jump across the tropopause:
\begin{equation}
    R(z^*) = 1 - \exp \bigg( \frac{-(z^* - 1)^2}{b_3} \bigg)
\end{equation}

We found that $b_1 = 5$, $b_2 = 0.5$, $b_3 = 0.01$ lead to a reasonable representation of the QBO and it's associated temperature anomalies (see Figure \ref{fa1}, right column). For instance, the meridional extent of the idealized QBO in the linear model corresponds well to the meridional extent of the real QBO, at least when compared to zonal winds estimates by ERA5 re-analysis from 1979-2020. While the vertical structure of the QBO is not exactly sinusoidal, the above parameters reasonably estimate the vertical wavelength of the observed QBO. Figure \ref{fa1}, left column, shows an example of the imposed QBO-like mean state in the stratosphere, using the above parameters and for $U = -0.5$.

\subsection*{Numerical Model}
The full mathematical description of the numerical system used in this study and modified from \citet{lin2021effect} is below:
\begin{align}
    \pder[u_0]{t} &= -ik \big[ \phi_{s} + V_1(p_t) s \big] + y v_0 - r u_0 \\ %+ \eta \dpder[u_0]{y} \\
    \pder[v_0]{t} &= \delta_x \Big[ -\pder[]{y} \big[ \phi_{s} + V_1(p_t) s \big] - y u_0 \Big] - r v_0 \\ %+ \eta \dpder[v_0]{y} \\
    \pder[u_1]{t} &= i k s + y v_1 - r u_1 \\ % + \eta \dpder[u_1]{y} \\
    \pder[v_1]{t} &= \delta_x \Big[ \pder[s]{y} - y u_1 \Big]  - r v_1 \\ %+ \eta \dpder[v_1]{y} \\
    \pder[s]{t} &= (1 + C) s_m + C_i q_v - w - \alpha (u_0 + u_1) - \chi s - r s \\ %+ \eta \dpder[s^*]{y} \\
    \gamma \pder[s_m]{t} &= -D s - \alpha (u_0 + u_1) - G w + C s_m + C_i q_v - r s_m \\ %+ \eta \dpder[s_m]{y}  \\
    w &= -ik(u_0 + u_1) - \pder[]{y} (v_0 + v_1)\\
    \pder[u_s]{t} &= -ik \phi_s + y v_s - r u_s  % + \eta \dpder[u_s]{y} \\
\end{align}
\begin{align}
    \pder[v_s]{t} &= \delta_x \Big[ -\pder[\phi_s]{y} - y u_s \Big] - r v_s \\ % + \eta \dpder[v_s]{y}  \\
    \pder[\phi_s]{t} &= - \int^{z}_\infty w^*_s S \: dz^* - r \phi_s \\ %+ \eta \dpder[\phi_s]{y} \\
    \rho_s w^*_s &= -B \Big( ik u_0 + \pder[v_0]{y} \Big) - \int_{z^* = 1}^{z} \Big[ \rho_s \Big( i k u_s(y, z^*) + \pder[]{y} v_s(y, z^*) \Big) \Big] dz^* \\
    \pder[q_v]{t} &= - U_c(z^* = z_c) \pder[q_v]{x} + \Upsilon w(z^* = z_c) \\
\end{align}
where all variables are defined in the with the exception of $r$, the sponge coefficient applied at the boundaries of the domain.

\subsection*{Cloud Radiative Feedbacks \label{sec_app_cloud}}
\subsubsection*{Observations}
\begin{figure*}
 \center
 \noindent\includegraphics[width=39pc,angle=0]{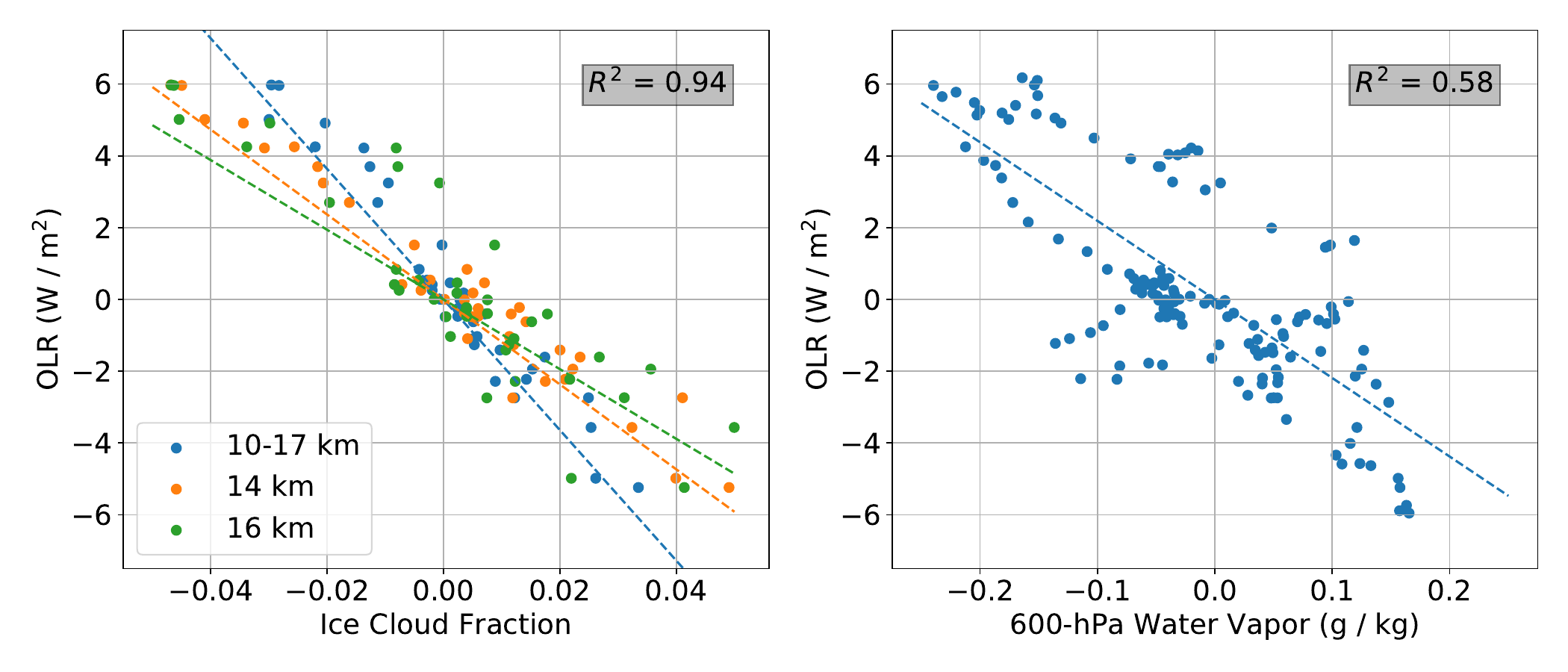}
 \caption{(Left) Ice cloud fraction against OLR across the Phase 2/-Phase 6 MJO monthly composite, where ice cloud fraction is (blue) averaged over 10-17 km, or taken at (orange) 14 km, and (green) 16 km. Linear regression lines are overlaid, while $R^2$ is shown for the 10-17 km average. (Right) Same as left, but for 600-hPa water vapor.} \label{fa0}
\end{figure*}

In this section, we show additional analysis of the observations of cirrus clouds and their associated radiative feedbacks. First, Figure \ref{fa0} shows the approximately linear relationship between OLR and both ice cloud fraction and 600-hPa water vapor, as deduced from MJO composites.

%% Figure A2 %%

\begin{figure}[h]
 \center
 \noindent \includegraphics[width=19pc,angle=0]{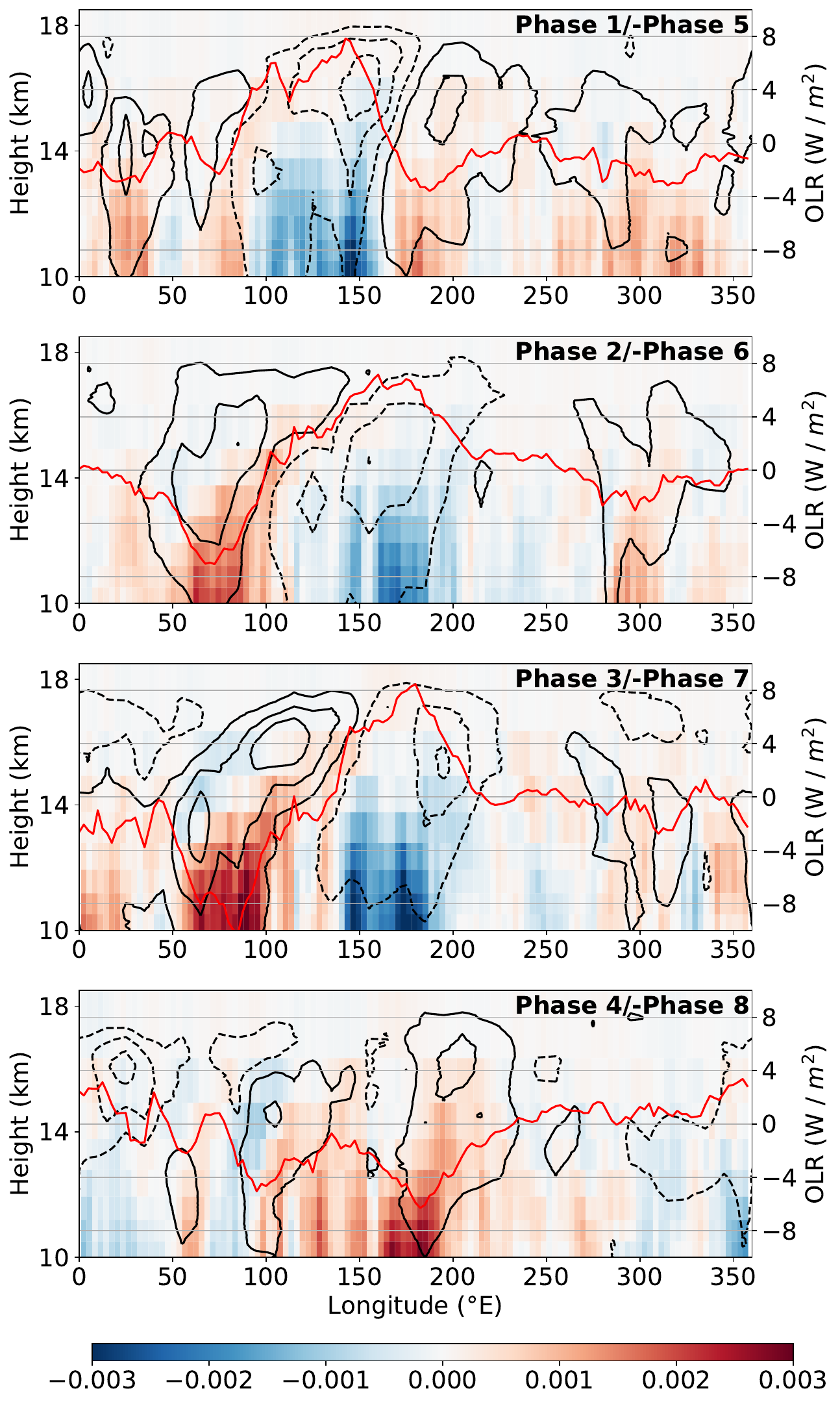}
 \caption{Same as Figure \ref{fig1} but with (contours) ice cloud fraction and (colors) vertical velocity (m/s), estimated from ERA5 re-analysis. Contour levels start at -0.07 with spacings of 0.02.} \label{fa3}
\end{figure}

Next, we present some results that show the validity of Equation (\ref{eq_qv}). Figure \ref{fa3}, which shows zonal-vertical MJO composites (similar to Figure 3 in the original manuscript) of vertical velocity and cloud fraction, indicates that in general, vertical velocity is positively correlated with ice cloud fraction anomalies. We have investigated this further by performing lead/lag linear regression between $w$ and ice cloud fraction.

%% Figure A3 %%

\begin{figure}[h]
 \center
 \noindent \includegraphics[width=19pc,angle=0]{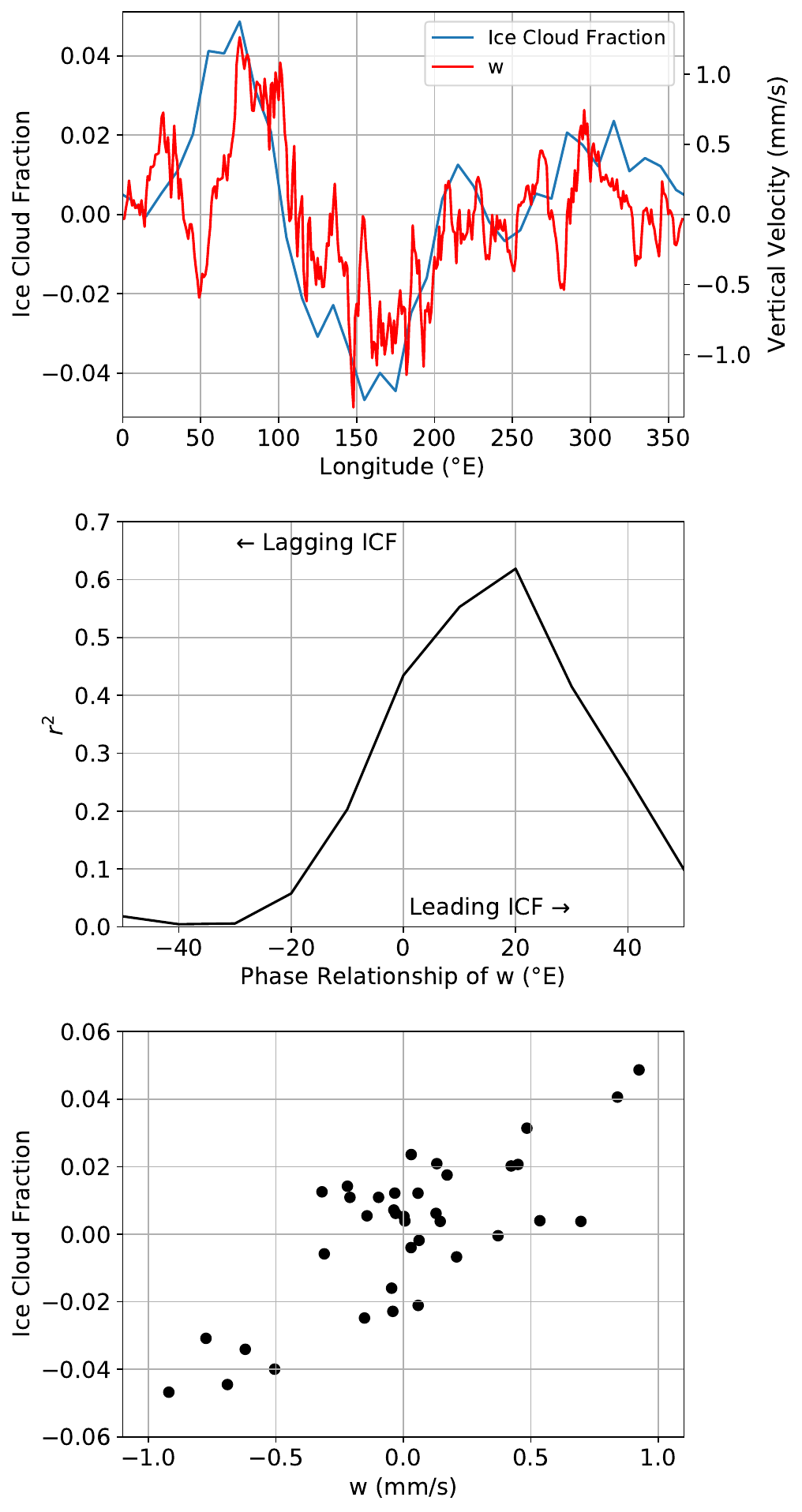}
 \caption{(Top) Longitudinal dependence of ice cloud fraction and vertical velocity, as estimated from ERA-5 re-analysis, at 14-km above sea level, in the Phase 2/-Phase 6 composite. (Middle) Coefficient of determination ($r^2$) for lagged linear regression of ice cloud fraction (ICF) on vertical velocity. Positive (negative) longitudinal phase displacements correspond to $w$ leading (lagging) ICF. (Bottom) Scatter relationship between $w$ and ICF at 14-km, for when $r^2$ is maximum in the lagged regression.} \label{fa4}
\end{figure}

Figure \ref{fa4}, top, shows the zonal dependence of the vertical velocity anomalies and ice cloud fraction anomalies at 14-km height in the Phase 2/-Phase 6 composite. There is a strong correlation between the two quantities. To analyze this more quantitatively, we perform lead/lag linear regression. Figure \ref{fa4}, middle, shows that the maximum correlation between $w$ and ICF (above 60\% variance explained) occurs when $w$ leads cloud fraction anomalies by around 20 degrees in longitude. This justifies the use of a prognostic equation for $q_v$, as in Equation (\ref{eq_qv}); in the linear model, if a mode is eastward propagating, the will be a phase lag between the prognostic variable ($q_v$) and the forcing term ($w$) that is proportional to the phase speed of the mode. We are qualitatively reproducing this using Equation (\ref{eq_qv}). Unfortunately, since the data are so sparse in space and time, there is not a good way to estimate the advective term in Equation (\ref{eq_qv}). However, the benefit of a simple linear model is that the potential impacts of zonal advection of cirrus clouds on the MJO can be easily explored. Finally, the scatter relationship between $w$ and ice-cloud fraction at 14-km, for when $w$ leads cloud fraction by 20 degrees in longitude, is shown in Figure \ref{fa4}, bottom. The relationship looks approximately linear, which justifies the linearization approach in Equation (\ref{eq_qv}).

%% Figure A4 %%

\subsubsection*{Linearization of Cloud Feedbacks}
For ice-only clouds, the CAM5 macrophysics parameterization is:
\begin{align*}
    \text{CF} &= \text{min} (1, \text{RH}_d^2) \\
    \text{RH}_d &= \text{max} \big[ 0, (RH_{ti} - RH_{imin}) / (RH_{imax} - RH_{imin}) \big] \\
    \text{RH}_{ti} &= \frac{q_v + q_i}{q^*_i}
\end{align*}
where CF is cloud fraction, $RH_{ti}$ is total ice water relative humidity, $RH_{\text{imin}}$ is the minimum relative humidity for ice (typically 0.8),
$RH_{\text{imax}}$ is the maximum relative humidity for ice (typically 1.1), $q_v$ is the water vapor mixing ratio, $q_i$ is the ice mass mixing ratio, and $q_i^*$ is the saturation vapor mixing ratio with respect to ice. Linearizing for $q_v$, ignoring changes to the ice mass (for simplicity), assuming that for the MJO, temperature anomalies are small compared to moisture anomalies \citep{ahmed2021quasi}, and non-dimensionalizing, we have:
\begin{equation}
    \text{CF}^\prime = \epsilon_i q_v \label{eqA_cf}
\end{equation}
where
\begin{equation}
    \epsilon_i = 2 \frac{\overline{\text{RH}}_{ti} - \text{RH}_{imin}}{(\text{RH}_{imax} - \text{RH}_{imin})^2}
\end{equation}
represents the production efficiency of ice clouds from water vapor anomalies. For $\overline{\text{RH}}_{ti} = 0.9$, $\text{RH}_{\text{imin}} = 0.8$, $\text{RH}_{\text{imax}} = 1.1$ \citep{gettelman2010global}, we have $\epsilon_i \approx 2$.

The modulation of tropospheric radiative cooling by both lower tropospheric water vapor and ice-clouds are incorporated using:
\[
    \dot{Q}^\prime = C s_m^\prime + C_i [\text{Ice Cloud Fraction}]^\prime
\]
where $s_m$ is the mid-level entropy anomaly (moisture deficit) that was originally formulated in \citet{khairoutdinov2018intraseasonal}. Since $q_v$ is a prognostic variable, the non-dimensional version of Equation (\ref{eqA_cf}) allows us to relate anomalies in water vapor to cloud fraction anomalies, where $C_i$ represents the strength of the ice-cloud radiative feedback.

%%%%%%%%%%%%%%%%%%%%%%%%%%%%%%%%%%%%%%%%%%%%%%%%%%%%%%%%%%%%%%%%%%%%%
% REFERENCES
%%%%%%%%%%%%%%%%%%%%%%%%%%%%%%%%%%%%%%%%%%%%%%%%%%%%%%%%%%%%%%%%%%%%%
% Make your BibTeX bibliography by using these commands:
\bibliographystyle{ametsocV6}
\bibliography{references}

\end{document}